\documentclass[%
reprint,
superscriptaddress,
 amsmath,amssymb,
 aps,
floatfix,
]{revtex4-1}

\usepackage{graphicx}
\usepackage{dcolumn}
\usepackage{bm}
\usepackage{multirow}
\usepackage{makecell}
\usepackage{array}
\usepackage[hyperindex,breaklinks]{hyperref}

\newcommand{\abs}[1]{\left| #1 \right|}

\begin{document}


\title{Chromatic Dynamics of an Electron Beam in a Plasma Based Accelerator}

\author{R. Ariniello}
\affiliation{Center for Integrated Plasma Studies, Department of Physics, University of Colorado Boulder, Boulder, Colorado 80309, USA}
\author{C. E. Doss}
\affiliation{Center for Integrated Plasma Studies, Department of Physics, University of Colorado Boulder, Boulder, Colorado 80309, USA}
\author{V. Lee}
\affiliation{Center for Integrated Plasma Studies, Department of Physics, University of Colorado Boulder, Boulder, Colorado 80309, USA}
\author{C. Hansel}
\affiliation{Center for Integrated Plasma Studies, Department of Physics, University of Colorado Boulder, Boulder, Colorado 80309, USA}
\author{J. R. Cary}
\affiliation{Center for Integrated Plasma Studies, Department of Physics, University of Colorado Boulder, Boulder, Colorado 80309, USA}
\affiliation{Tech-X Corporation, Boulder, Colorado 80301, USA}
\author{M. D. Litos}
\affiliation{Center for Integrated Plasma Studies, Department of Physics, University of Colorado Boulder, Boulder, Colorado 80309, USA}

\date{\today}

\begin{abstract}
We present a theoretical investigation of the chromatic dynamics of the witness beam within a plasma based accelerator. We derive the single particle motion of an electron in an ion column within a nonlinear, blowout wake including adiabatic dampening and adiabatic variations in plasma density. Using this, we calculate the evolution of the beam moments and emittance for an electron beam. Our model can handle near arbitrary longitudinal phase space distributions. We include the effects of energy change in the beam, imperfect wake loading, initial transverse offsets of the beam, and mismatch between the beam and plasma. We use our model to derive analytic saturation lengths for the projected, longitudinal slice, and energy slice emittance under different beam loading conditions. Further, we show that the centroid oscillations and spot sizes vary between the slices and the variation depends strongly on the beam loading. Next, we show how a beam evolves in a full plasma source with density ramps and show that the integral of the plasma density along the ramp determines the impact on the beam. Finally, we derive several simple scaling laws that show how to design a plasma based injector to produce a target beam energy and energy spread.
\end{abstract}


\maketitle

\section{\label{sec:intro}Introduction}

Accelerators, in the form of high energy colliders and light sources, have proven to be important tools for a diverse range of research fields. Unfortunately, the size and cost of these machines are prohibitive, especially at the energy frontier. Plasma based accelerators are promising, compact alternatives that have been shown to produce accelerating gradients two to three orders of magnitude larger than conventional radio frequency accelerators. Great progress has been made in demonstrating low energy spread beams and high efficiency acceleration \cite{Blumenfeld2007, Litos2014, Litos2016}. Colliders and light sources, however, require beams with high brightness and thus place strict limits on the beam's transverse emittance. Current plasma based accelerators struggle to meet these strict requirements. The emittance of the accelerated beam, called the witness beam, typically grows considerably as the beam traverses an accelerating stage. 

Multiple mechanisms contribute to the beam's emittance growth within the plasma stage. Recent work has primarily focused on emittance growth due to mismatch between the plasma and the beam \cite{Michel2006, Mehrling2012, Floettmann2014, Dornmair2015, Xu2016, Aschikhin2018, ariniello:2019prab, Zhao2020} and beam instabilities \cite{Whittum1991, Lampe1993, Geraci2000b, Deng2006b, Huang2007, Mehrling2017, Lebedev2017, Mehrling2018, Mehrling2018a, Mehrling2019}. Yet the emittance growth of a transversely offset beam---due to chromatic phase spread---has only been briefly considered: Refs.~\cite{Assmann1998, Lindstrom2016b} derived simple expressions for the saturated emittance and initial growth rate, Ref.~\cite{Thevenet2019} considered emittance in the presence of a laser driver, and Ref.~\cite{Raubenheimer2000} worked out the saturated emittance for an injection mismatch in a conventional accelerator. In the case of mismatch, emittance growth due to energy gain in the presence of plasma ramps, or with the inclusion of wake loading, has not been considered. Typically, the longitudinal phase space is assumed to take on a simple form and previous approaches cannot handle arbitrary distributions. Further, the longitudinal slice emittance has only been briefly investigated \cite{Xu2017, Thevenet2019, Dalichaouch2020}, while the energy slice emittance has yet to receive serious attention despite its importance in a transverse gradient undulator \cite{Huang2012,Smith1979,Baxevanis2014}. 

We derive the projected emittance, slice emittance (longitudinal and energy), and moment evolution of an electron beam travelling in a nonlinear, blowout wake including the effects of adiabatic plasma ramps, energy gain, beam loading, initial transverse offsets, and the initial longitudinal phase space distribution. We start by deriving the single particle motion of an electron, including energy change, in an ion column with adiabatically varying density. Next, we derive the evolution of the beam moments and the emittance in a way that allows straightforward evaluation of the slice and projected beam parameters. We separate the effect of the beam's initial longitudinal phase space into a single parameter that can be analytically evaluated in simple cases. For complicated phase space distributions, this parameter can be evaluated numerically while retaining the rest of the analytic formulation.

To show the broad applicability of our model, we present several examples. First, we consider a beam in a uniform plasma with continuous energy gain. We use two simple models for the longitudinal accelerating field to represent a beam that overloads the wake and one that does not sufficiently load the wake. In both cases, we derive analytic saturation lengths for the projected, longitudinal slice, and energy slice emittance. We show that beam loading causes particles to mix between energy slices leading to growth in the energy slice emittance. We then calculate the transverse offset of the different longitudinal slices and the spot size of the different energy slices. We show that the slice parameters depend on the beam loading - a result with experimental consequences. Second, we calculate the witness beam evolution through a full plasma source with density ramps, and show that the integral over the ramp density determines how much of an impact the ramp has on the beam. We show why it is the ramp shape, and not the length, that is important. Third, we calculate the beam evolution in a plasma based injector and derive simple, analytic expressions for designing an injector to produce a beam with a target energy and energy spread. These examples demonstrate the generality of our approach. It combines multiple effects in a straightforward analytic framework.

\section{Single Particle Motion}

In a plasma based accelerator operating in the blowout regime, all of the plasma electrons are evacuated from the center of the wake leaving a column of ions behind. If the plasma has a transversely uniform density and ion motion is neglected, the ions produce a linear focusing force on the witness beam that is independent of $\xi=ct-s$, where $s$ is the distance along the accelerator \cite{Lu2006, Lu2006a}. The equations of motion in the transverse directions for an electron in the witness beam are decoupled and given by \cite{Xu2014a, Aschikhin2018}
\begin{equation} \label{eqn:EOM}
\frac{d^2x}{ds^2} +\frac{\gamma_b'}{\gamma_b}\frac{dx}{ds}+ K(s)x = 0,
\end{equation}
where a prime denotes a derivative with respect to $s$. The focusing strength is given by
\begin{equation}
K(s) = \frac{\omega_p^2(s)}{2\gamma_b(s)c^2}.
\end{equation}
In general, $K$ is a function of $s$ through the local plasma density and the particle energy. The plasma frequency $\omega_p$ is defined as $\omega_p^2(s)=n(s)e^2/(m_e\epsilon_0)$, $\gamma_{b}(s)$ is the relativistic factor of the electron, $c$ is the speed of light, $n(s)$ is the plasma density, $e$ is the elementary charge, $m_e$ is the mass of the electron, and $\epsilon_0$ is the permittivity of free space. The betatron wavenumber is defined as $k_\beta^2(s)=K(s)$.

An approximate solution to the equation of motion can be derived for a plasma density that varies adiabatically in $s$. The adiabatic condition is defined as \cite{ariniello:2019prab}:
\begin{equation} \label{eq:Aparam}
\left|\alpha_m(s)\right|\ll1,
\end{equation}
where $\alpha_m$ is one of the matched Courant-Snyder (CS) parameters defined for a single particle as: $\beta_m=1/k_\beta$, $\alpha_m=-\beta'_m/2$, and $\gamma_m=(1+\alpha_m^2)/\beta_m$. $\beta_m$ sets the natural length scale of the transverse evolution. In a uniform plasma with no energy gain, $\beta_m$ is constant and $\alpha_m=0$. The matched CS parameters are functions of $s$ due to the variation in plasma density and particle energy along the accelerator. 

In the absence of energy gain, the transverse motion of a single particle in an adiabatically varying plasma density is given by \cite{ariniello:2019prab, Zhao2020}
\begin{equation} \label{eqn:uniformspm}
\begin{split}
x = &x_0\sqrt{\frac{\beta_m}{\beta_{m0}}}(\cos\phi+\alpha_{m0}\sin\phi) + x'_0\sqrt{\beta_m\beta_{m0}}\sin\phi\\
x' = &x_0\frac{(\alpha_{m0}-\alpha_m)\cos\phi-(1+\alpha_{m0}\alpha_m)\sin\phi}{\sqrt{\beta_m\beta_{m0}}} \\
&+x'_0\sqrt{\frac{\beta_{m0}}{\beta_m}}(\cos\phi-\alpha_{m0}\sin\phi),
\end{split}
\end{equation}
where the subscript 0 indicates the initial value of a variable at $s=s_0$ and $x'=dx/ds=p_x/p_s$, $p_x$ and $p_s$ are the longitudinal and transverse momentum of the particle, respectively. The betatron phase advance is defined as
\begin{equation} \label{eqn:phi}
\phi=\int_{s_0}^sk_\beta(s')ds'.
\end{equation}

We are interested in the general case where the particles can gain and lose energy. Energy gain (loss) results in a reduction (increase) in the amplitude of the oscillations due to adiabatic dampening. We use the ansatz that the single particle motion has an additional $s$ dependent amplitude $A(s)$:
\begin{equation}
\begin{split}
x = A(s)x_c = &x_0A(s)\sqrt{\frac{\beta_m}{\beta_{m0}}}(\cos\phi+\alpha_{m0}\sin\phi) \\
&+ x'_0A(s)\sqrt{\beta_m\beta_{m0}}\sin\phi,
\end{split}
\end{equation}
where $x_c$ is the position of a single particle with constant energy given by Eq.~(\ref{eqn:uniformspm}). $x_c$ satisfies the differential equation $x_c''+k_\beta^2x_c=0$. Inserting the ansatz into Eq.~(\ref{eqn:EOM}) and requiring that $A$ can take on any value gives a differential equation for $A$:
\begin{equation}
2A'+\frac{\gamma_b'}{\gamma_b}A=0.
\end{equation}
The solution of which is $A(s)=\sqrt{\gamma_{b0}/\gamma_b}$. For $Ax_c$ to approximately satisfy the equation of motion, Eq.~(\ref{eqn:EOM}), the relative change in energy over one betatron period must be small $\gamma_b'/\gamma_b\ll k_\beta$ and $\gamma_b''/\gamma_b\ll k_\beta^2$. 

The transport matrix $M$ defines the motion of the particle based on its initial conditions
\begin{equation} \label{eqn:matrixEq}
\begin{pmatrix} x\\ x' \end{pmatrix} = M \begin{pmatrix} x_0\\ x_0' \end{pmatrix} = 
\begin{pmatrix} M_{11} & M_{12} \\ M_{21} & M_{22} \end{pmatrix} \begin{pmatrix} x_0\\ x_0' \end{pmatrix}.
\end{equation}
Combining the adiabatic dampening term $A(s)$ with $x_c$ gives the transport matrix for a particle in an adiabatic plasma with energy change:
\begin{equation} \label{eqn:spm}
\begin{split}
M_{11}&= \sqrt{\frac{\gamma_{b0}}{\gamma_b}\frac{\beta_m}{\beta_{m0}}}(\cos\phi+\alpha_{m0}\sin\phi) \\
M_{12}&= \sqrt{\frac{\gamma_{b0}}{\gamma_b}\beta_m\beta_{m0}}\sin\phi \\
M_{21}&= \sqrt{\frac{\gamma_{b0}}{\gamma_b}}\frac{(\alpha_{m0}-\alpha_m)\cos\phi-(1+\alpha_{m0}\alpha_m)\sin\phi}{\sqrt{\beta_m\beta_{m0}}}\\
M_{22}&= \sqrt{\frac{\gamma_{b0}}{\gamma_b}\frac{\beta_{m0}}{\beta_m}}(\cos\phi-\alpha_{m0}\sin\phi).
\end{split}
\end{equation}
In a uniform plasma, $\beta_m=\beta_{m0}\sqrt{\gamma_b/\gamma_{b0}}$. Inserting $\beta_m$ into Eq.~(\ref{eqn:spm}) and assuming the energy gain over a single betatron period is small, $\alpha_m\approx0$, we recover the expression for single particle motion in a uniform plasma from \cite{Xu2014a, Aschikhin2018}. 

\begin{figure}[bt]
  \centering
  \includegraphics[width=3.37in]{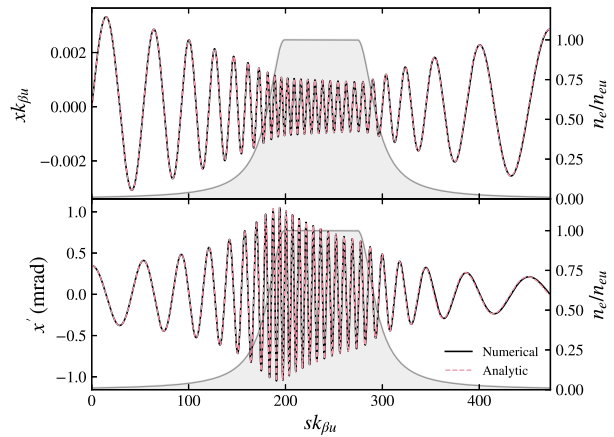}
  \caption{Single particle motion in a plasma accelerator with adiabatic ramps. The particles energy doubles within the accelerator. The top plot shows the evolution of the particle's transverse position $x$. The bottom plot shows the evolution of the particle's trajectory angle $x'$. The solid line shows the numerical solution to the transverse equation of motion, the dashed line shows the analytic expression from Eq.~(\ref{eqn:spm}). The longitudinal plasma density is shown by the shaded region. \label{fig:spm}}
\end{figure}

Fig.~\ref{fig:spm} shows a comparison between Eq.~(\ref{eqn:matrixEq}) and a numerical solution to Eq.~(\ref{eqn:EOM}) for a single electron travelling through a plasma based accelerator. The plasma stage is designed to double the particle's energy and has identical adiabatic density ramps on each end. The entrance ramp reduces the incoming beta function by a factor of 10, and the beam undergoes approximately 12 betatron periods within the uniform plasma section. $k_{\beta u}$ is the betatron wavenumber at the start of the uniform plasma section, located at $sk_{\beta u}=200$. The parameters used here are similar to those found in current beam driven plasma wakefield accelerator experiments such as those at FACET-II \cite{Joshi2018,Yakimenko2019} and FLASHForward \cite{Aschikhin2016,Lindstrom2021a}. The analytic solution shows excellent agreement with the numerical solution; it accurately captures both the adiabatic dampening and focusing in the ramp. In the next section, we use this solution to derive general expressions for the evolution of the beam's moments and emittance.

\section{\label{sec:emt}Evolution of the Beam Moments and Projected Emittance}

The transverse beam quality can be quantified using the normalized emittance \cite{Floettmann2003}: 
\begin{equation}
\epsilon_n=(1/m_ec)\sqrt{\sigma_x^2\sigma_{p_x}^2-\sigma_{xp_x}^2},
\end{equation}
where the beam sizes are $\sigma_x^2=\left<x^2\right>-\left<x\right>^2$, $\sigma_{p_x}^2=\left<p_x^2\right>-\left<p_x\right>^2$ and the correlation is $\sigma_{xp_x}=\left<xp_x\right>-\left<x\right>\left<p_x\right>$. In the ultra-relativistic limit the normalized emittance simplifies to
\begin{equation} \label{eq:normEmit}
\epsilon_n^2=\left<\gamma_b\right>^2\left(\sigma_\delta^2\sigma_x^2\left<x'^2\right>+\epsilon^2\right),
\end{equation}
where $\sigma_\delta^2=(\left<\gamma_b^2\right>-\left<\gamma_b\right>^2)/\left<\gamma_b\right>^2$ and $\epsilon$ is the geometric emittance defined as 
\begin{equation}
\epsilon^2=\sigma_x^2\sigma_{x'}^2-\sigma_{xx'}^2
\end{equation}
with $\sigma_{x'}^2=\left<x'^2\right>-\left<x'\right>^2$ and $\sigma_{xx'}=\left<xx'\right>-\left<x\right>\left<x'\right>$. In most cases, the first term in Eq.~(\ref{eq:normEmit}) is very small and the geometric emittance dominates, giving the more familiar formula $\epsilon_n\approx\left<\gamma_b\right>\epsilon$.

\begin{figure}[bt]
  \centering
  \includegraphics[width=3.37in]{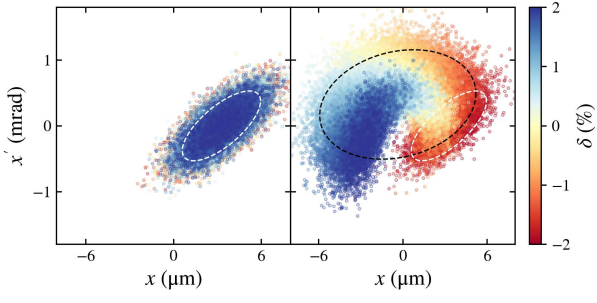}
  \caption{When the witness beam is mismatched or offset transversely with respect to the ion column, it will undergo emittance growth through chromatic dephasing. Different energy slices of the beam will rotate at different frequencies in transverse phase space, leading to growth in the projected emittance (the area enclosed by the white dashed line grows to the area enclosed by the black dashed line).  \label{fig:chromEm}}
\end{figure}

Emittance growth results from the $\gamma_b$ dependence of $k_\beta$. Different energy slices of the beam oscillate with different frequencies in the ion channel. Over time the different slices dephase and no longer overlap in transverse phase space as shown in Fig.~\ref{fig:chromEm}. Even if the beam has no initial energy spread, this chromatic dephasing will occur if imperfect wake loading induces energy spread in the beam during acceleration. This same effect dampens out the centroid oscillations of an offset beam. To calculate the growth explicitly requires the evaluation of the moments of the beam distribution.

The beam moments can be evaluated under the assumption that there is no initial correlation between the longitudinal distribution and the transverse distribution. In this case the phase space distribution of the witness beam at $s_0$ can be written as $f_0(x, x', \xi, \delta) =  f_\perp(x, x')f(\xi, \delta)$, where $\delta=(p-p_0)/p_0$ parameterizes the energy spread in the beam and $p_0$ is the momentum of the reference particle. Here, we are ignoring the vertical transverse dimension because it is decoupled and evolves independently in an analogous fashion. We let $f_\perp$ describe the witness beam without offset; i.e., the beam centroid is located at the origin in transverse phase space: $\int dxdx'\,xf_\perp(x, x')=0$ and $\int dxdx'\,x'f_\perp(x, x')=0$. Further, we assume that the distribution is normalized to 1. 

If the witness beam is then given an initial offset of $\Delta x$ in $x$ and $\Delta x'$ in $x'$, the beam distribution at $s_0$ can be written as $f_\Delta(x, x', \xi, \delta) = f_\perp(x-\Delta x, x'-\Delta x')f(\xi, \delta)$. Single particle evolution is described by Eq.~(\ref{eqn:matrixEq}) and is a function of $s$, $x_0$, $x'_0$, $\xi_0$, $\delta_0$: $x(s, x_0, x'_0, \xi_0, \delta_0)$, where the $\xi_0$ and $\delta_0$ dependence enters through the relativistic factor $\gamma_b(s,\xi_0,\delta_0)$. The functional form of $\gamma_b$ depends on the acceleration model chosen, we show several in the examples. Making the change of variables to $u=x-\Delta x$ and $u'=x'-\Delta x'$, we can write the single particle evolution in terms of $u$ and $u'$: $x(s, u_0+\Delta x, u'_0+\Delta x', \xi_0, \delta_0)$. The first beam moment is given by:
\begin{equation} \label{eq:uupint}
\left<x\right> = \int d\xi_0d\delta_0f(\xi_0, \delta_0)\int du_0du_0'\,x(u_0, u'_0)f_\perp(u_0, u_0').
\end{equation}
The other beam moments, $\left<x^2\right>$, $\left<x'\right>$, $\left<x'^2\right>$, and $\left<xx'\right>$ take the same form. 

To make the integral tractable, we make the assumption that the energy spread is small: $\delta\ll1$. This allows us to treat the matched CS parameters and the adiabatic dampening term, $A=\sqrt{\gamma_{b0}/\gamma_b}$, in Eq.~(\ref{eqn:spm}) as constant with respect to $\xi_0$ and $\delta_0$. We use bars to denote values for the reference particle: $\bar{\gamma}_b=\gamma_b(s, \xi_0=0, \delta_0=0)$. Using $\gamma_b=\bar{\gamma}_b$ in $A$, $\beta_m$, $\alpha_m$, and $\gamma_m$ for all particles defines a matched set of CS parameters for the beam; this is in contrast to the approach taken in Ref.~\cite{ariniello:2019prab}, where different matched CS parameters are defined for each energy slice. The only dependence on $\xi_0$ and $\delta_0$ remaining in Eq.~(\ref{eqn:spm}) is in the energy dependence of the betatron phase $\phi$.

We have to integrate over $u$ and $u'$ in Eq.~(\ref{eq:uupint}) first because $x$ depends on the longitudinal coordinates. The $u$ and $u'$ integrals can be written in terms of moments of the witness beam's initial (non-offset) phase space distribution. The second central moments of the phase space distribution, $\sigma_x$, $\sigma_{x'}$, and $\sigma_{xx'}$, are defined as: $\sigma_x^2=\int dxdx'\,x^2f_\perp(x, x')$, $\sigma_{x'}^2=\int dxdx'\,x'^2f_\perp(x, x')$, and $\sigma_{xx'}=\int dxdx'\,xx'f_\perp(x, x')$. These moments can be expressed in terms of the CS parameters: $\beta=\sigma_x^2/\epsilon$, $\gamma=\sigma_{x'}^2/\epsilon$, and $\alpha=-\sigma_{xx'}/\epsilon$. 

Evaluating the $u$, $u'$ integrals, and writing the beam moments in terms of the beam's initial CS parameters gives the following expressions for the moments of the offset beam:
\begingroup
\allowdisplaybreaks
\begin{align} \label{eqn:moments}
\begin{split}
\left<x\right> =& \sqrt{\frac{\bar{\gamma}_{b0}}{\bar{\gamma}_b}\beta_m\epsilon_0}(d_1C_1+d_2S_1)
\end{split}
\\[2ex]
\begin{split}
\left<x^2\right> =& \frac{\bar{\gamma}_{b0}}{\bar{\gamma}_b}\beta_m\epsilon_0(a+b_1C_2+b_2S_2)
\end{split}
\\[2ex]
\begin{split}
\left<x'\right> =& \sqrt{\frac{\bar{\gamma}_{b0}}{\bar{\gamma}_b}\frac{\epsilon_0}{\beta_m}}[-d_1S_1+d_2C_1-\alpha_m(d_1C_1+d_2S_1)]
\end{split}
\\[2ex]
\begin{split}
\left<x'^2\right> =& \frac{\bar{\gamma}_{b0}}{\bar{\gamma}_b}\frac{\epsilon_0}{\beta_m}[(1+\alpha_m^2)a-(1-\alpha_m^2)(b_1C_2+b_2S_2) \\
&+2\alpha_m(b_1S_2-b_2C_2)]
\end{split}
\\[2ex]
\begin{split} \label{eqn:moments_end}
\left<xx'\right> =& \frac{\bar{\gamma}_{b0}}{\bar{\gamma}_b}\epsilon_0[b_2C_2-b_1S_2-\alpha_m(a+b_1C_2+b_2S_2)],
\end{split}
\end{align}
\endgroup
where we have kept terms of order $\alpha_{m}^2$. The $a$, $b$, and $d$ terms are all unit-less constants that depend on the beam's initial conditions:
\begingroup
\allowdisplaybreaks
\begin{align*}
d_1 =& \frac{\Delta x}{\sqrt{\beta_{m0}\epsilon_0}} \\
d_2 =& \alpha_{m0}\frac{\Delta x}{\sqrt{\beta_{m0}\epsilon_0}}+\sqrt{\frac{\beta_{m0}}{\epsilon_0}}\Delta x' \\
a =& \frac{1}{2}\left(\beta_0\gamma_{m0}+\gamma_0\beta_{m0}-2\alpha_0\alpha_{m0}+d_1^2+d_2^2\right) \\
b_1 =& \frac{\beta_0}{\beta_{m0}}+d_1^2-a \\
b_2 =& d_1d_2-\alpha_0+\alpha_{m0}\frac{\beta_0}{\beta_{m0}}.
\end{align*}
\endgroup
The first three constants have physical meaning; $d_1$ and $d_2$ are the normalized phase space coordinates for the motion of the beam centroid and, as we will show later, $a$ is the relative emittance growth at saturation.

The $s$ dependence of the moments enters through $\bar{\gamma}_b$ directly (adiabatic dampening) and through the matched CS parameters (plasma focusing), which are functions of $\bar{\gamma}_b$; however, the primary dependence on $s$ is given by the $C$ and $S$ terms which capture both the chromatic phase spreading and the betatron oscillations. These terms are integrals over the longitudinal phase space:
\begingroup
\allowdisplaybreaks
\begin{align*}
C_1 =& \int d\xi_0d\delta_0f(\xi_0, \delta_0)\cos[\phi(\xi_0, \delta_0)] \\
S_1 =& \int d\xi_0d\delta_0f(\xi_0, \delta_0)\sin[\phi(\xi_0, \delta_0)] \\
C_2 =& \int d\xi_0d\delta_0f(\xi_0, \delta_0)\cos[2\phi(\xi_0, \delta_0)] \\
S_2 =& \int d\xi_0d\delta_0f(\xi_0, \delta_0)\sin[2\phi(\xi_0, \delta_0)].
\end{align*}
\endgroup
The two dimensional longitudinal phase space distribution can always be reduced to a one dimensional distribution of betatron phase advance $f_\phi(\phi)$. The integrals can then be written in the form:
\begin{equation}
C_1 = \int d\phi\,f_{\phi}(\phi)\cos\phi,
\end{equation}
where we have only written the $C_1$ integral for brevity. The $s$ dependence enters through the variation of the $f_\phi$ distribution as the beam propagates through the plasma. At this point, the saturated emittance can be determined. As the different energy components of the beam dephase, the width of the $f_\phi$ distribution will grow. At saturation, the width is much larger than $2\pi$ and the $C$ and $S$ integrals tend towards zero. Setting $C$ and $S$ to zero in the moments and calculating the emittance gives the saturated emittance
\begin{equation} \label{eqn:sat}
\begin{split}
\epsilon_{nsat} =&\epsilon_{n0}\sqrt{1+\sigma_\delta^2}\frac{1}{2}\bigg(\beta_0\gamma_{m0}+\gamma_0\beta_{m0}-2\alpha_0\alpha_{m0} \\
&+\frac{\Delta x^2}{\beta_{m0}\epsilon_0}+\Delta x'^2\frac{\beta_{m0}}{\epsilon_0}+\Delta x\Delta x'\frac{\alpha_{m0}}{\epsilon_0}\bigg),
\end{split}
\end{equation}
where terms higher than $\alpha_m$ have been dropped. The pre-factor $\sqrt{1+\sigma_\delta^2}$ can be set equal to one if the final energy spread is sufficiently small. 

The detailed emittance evolution can be found by rewriting the $C$ and $S$ integrals. We start by writing the betatron phase as $\phi=\bar{\phi}+\Delta\phi(\xi_0, \delta_0)$. Further, we note that the four $C$ and $S$ integrals can be written as the real and imaginary parts of two complex integrals. After removing the $e^{i\bar{\phi}}$ term from the them, the integrals generically evaluate to a pair of complex numbers with real amplitudes $H_1$, $H_2$, and arguments $\psi_1$, $\psi_2$:
\begingroup
\allowdisplaybreaks
\begin{align}
I_1&=\int d\xi_0 d\delta_0f(\xi_0, \delta_0)e^{i\Delta\phi(\xi_0, \delta_0)}=H_1e^{i\psi_1}, \label{eqn:I1}\\
I_2&=\int d\xi_0 d\delta_0f(\xi_0, \delta_0)e^{i2\Delta\phi(\xi_0, \delta_0)}=H_2e^{i\psi_2}. \label{eqn:I2}
\end{align}
\endgroup
The $C$ and $S$ integrals are then given by
\begingroup
\allowdisplaybreaks
\begin{align*}
C_1 =& H_1\cos(\bar{\phi}+\psi_1) \\
S_1 =& H_1\sin(\bar{\phi}+\psi_1) \\
C_2 =& H_2\cos(2\bar{\phi}+\psi_2) \\
S_2 =& H_2\sin(2\bar{\phi}+\psi_2).
\end{align*}
\endgroup
If one of three conditions is met or approximately met, the expression for the emittance simplifies significantly. The first condition is if $\psi_1=\psi_2=0$. The second is if $\psi_1=\psi_2=\pi/2$. The third is if $\psi_2=2\psi_1$. In all the examples we consider later, either the first or third condition is satisfied. The emittance then evolves according to
\begin{widetext}
\begin{equation} \label{eq:emGrowth}
\epsilon_n=\epsilon_{nsat}\sqrt{1-\frac{b_1^2+b_2^2}{a^2}H_2^2
-\frac{1}{a^2}\left\{a\left(d_1^2+d_2^2\right)
-\left[b_1\left(d_1^2-d_2^2\right)
+2b_2d_1d_2\right]H_2\right\}H_1^2},
\end{equation}
\end{widetext}
where $H_2$ terms describe emittance growth due to mismatch and $H_1$ terms describe emittance growth due to transverse offsets in position and angle. In addition, $H_1$, when combined with the adiabatic dampening, describes the dampening of the beams centroid oscillations. The $H(s)$ functions describe the amount of chromatic coherence in the beam, they get smaller as the beam propagates and dephases. $\psi_1$ is the betatron phase difference between the projected beam and the reference particle.

\begin{figure}[bt]
  \centering
  \includegraphics[width=3.37in]{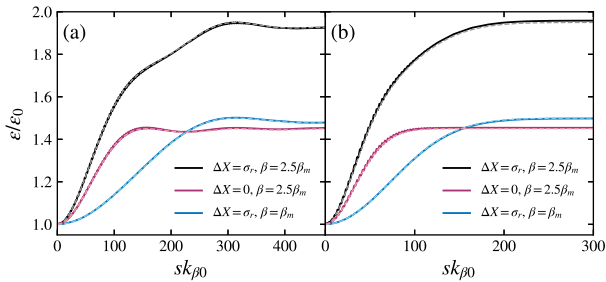}
  \caption{Emittance growth of a beam in a uniform plasma without energy gain. The beam is transversely offset in the blue and black curves. The beam is mismatched in the red and black curves. (a) shows the emittance evolution for a beam with uniform energy spread, (b) shows the evolution for a beam with Gaussian energy spread. The dashed lines represent the analytic theory and solid lines the numerical particle tracking results. In all cases, the energy spread is 2\%. The mismatch driven emittance growth saturates in half the distance of the offset driven emittance growth. The final saturated emittance growth is the sum of the mismatch and offset. \label{fig:uniform}}
\end{figure}

In Fig.~\ref{fig:uniform} we compare the results of numerical particle tracking with Eq.~(\ref{eq:emGrowth}) for a uniform plasma without energy gain. The witness beam has an initial energy spread and is either mismatch to the plasma (red), offset transversely from the drive beam (blue), or both mismatched and offset (black). In this simple case, Eq.~(\ref{eqn:phi}) gives the betatron phase advance as 
\begin{equation*}
\phi=\bar{\phi}\frac{1}{\sqrt{1+\delta_0}}.
\end{equation*}
Here, $\phi$ is independent of $\xi_0$ because the accelerating field is constant (in this case zero) in $\xi_0$. The betatron phase advance of the reference particle is $\bar{\phi}=\bar{k}_\beta s$, where $\bar{k}_\beta=\omega_p/(c\sqrt{2\bar{\gamma}_b})$ is the betatron wavenumber of the reference particle. Expanding to first order in $\delta_0$ simplifies the expression to $\phi=\bar{\phi}-\bar{\phi}\delta_0/2$. Since $\phi$ depends linearly on $\delta_0$, the $I_1$ and $I_2$ integrals are Fourier transforms of the energy distribution with frequencies $\omega_\delta=\bar{\phi}/2$ and $2\omega_\delta=\bar{\phi}$, respectively. If the distribution of energy is even (symmetric about $\bar{\gamma}_b$), then $H_1=\hat{f}_\delta(\omega_\delta)=\hat{f}_\delta(\bar{\phi}/2)$, $H_2=\hat{f}_\delta(\bar{\phi})$, and $\psi_1=\psi_2=0$. Here, $\hat{f}_\delta$ is the Fourier transform of the energy distribution $f_\delta$. Eq.~(\ref{eq:emGrowth}) then fully describes the evolution of the projected emittance.

In the case of a uniform energy spread between $-\Delta\delta/2$ and $\Delta\delta/2$, $\hat{f}_\delta(\bar{\phi})=\mathrm{sinc}\Delta\Phi$ and $\hat{f}_\delta(\bar{\phi}/2)=\mathrm{sinc}(\Delta\Phi/2)$, where $\Delta\Phi$ is the range of $\phi$ spanned by the energy spread $\Delta\Phi=\Delta\delta\bar{\phi}$. In the absence of transverse offset, this formula reduces to that given in Ref.~\cite{Xu2016}. The emittance growth of such a beam is shown in Fig.~\ref{fig:uniform}(a) for various initial mismatches and transverse offsets. The analytic expression shows excellent agreement with the numerical particle tracking code. The small oscillations of the numerical result about the theoretical emittance from Eq.~(\ref{eq:emGrowth}) are due to treating $\beta_m$ and $\gamma_b$ as constants and are of order $\sigma_\delta^2$ (for more details see Ref.~\cite{ariniello:2019prab}). 

There are two length scales for the emittance growth. The growth due to mismatch saturates when $\Delta\Phi=\pi$ giving a saturation length of $s_2=\pi/(\Delta\delta\bar{k}_\beta)$, while the emittance growth due to transverse offset requires twice the length to saturate: $s_1=2\pi/(\Delta\delta\bar{k}_\beta)$. 

In the case of a Gaussian distribution of energy, $f_\delta(\delta_0)=\frac{1}{\sqrt{2\pi}}\exp\left(-\frac{\delta_0^2}{2\sigma_\delta^2}\right)$ and $\hat{f}_\delta(\bar{\phi})=\exp\left(-\frac{\bar{\phi}^2\sigma_\delta^2}{2}\right)$. Without any transverse offset, $\Delta x =\Delta x'=0$, the solution reduces to that given in Ref.~\cite{Aschikhin2018}. The theoretical solution is compared to numerical particle tracking in Fig.~\ref{fig:uniform}(b). As can be seen in the figure, the saturated emittance is the sum of the emittance growth due to an offset and mismatch. 

\section{\label{sec:slice}Evolution of the Slice Moments and Slice Emittance}
Although the projected emittance is normally taken as the figure of merit, both the longitudinal slice emittance and the energy slice emittance are of practical interest. We start with the longitudinal slice emittance, which is of importance for both light sources and colliders. For an individual slice at $\xi$, the $I_1$ integral takes the form:
\begin{equation} \label{eqn:longslice}
\begin{split}
I_1(\xi)&=\frac{1}{N(\xi)}\int d\xi_0 d\delta_0 f(\xi, \delta_0)\delta_D(\xi_0-\xi)e^{i\Delta\phi(\xi_0,\delta_0)} \\
I_1(\xi)&=\frac{1}{N(\xi)}\int d\delta_0 f(\xi, \delta_0)e^{i\Delta\phi(\xi,\delta_0)}=H_1(\xi)e^{i\psi_1(\xi)},
\end{split}
\end{equation}
where $N(\xi)=\int d\delta_0 f(\xi, \delta_0)$ is a normalization factor and $\delta_D$ is the Dirac delta function. $I_2$ has the same form with $\Delta\phi$ replaced by $2\Delta\phi$. The expressions for the moments and emittance of each slice are given by Eqs.~(\ref{eqn:moments})-(\ref{eqn:moments_end}) and Eq.~(\ref{eq:emGrowth}), but with $H$ and $\psi$ replaced by the $\xi$ dependent expressions from Eq.~(\ref{eqn:longslice}). Further accuracy can be achieved by using $\bar{\gamma_b}$ and $\beta_m$ calculated for each slice rather than the beam as a whole.

Similarly, the energy slice emittance is found by replacing the $I_1$ and $I_2$ integrals with
\begin{equation} \label{eqn:deltaslice}
\begin{split}
I_1&(\delta)= H_1(\delta)e^{i\psi_1(\delta)}=\\
&\frac{1}{N(\delta)}\int d\xi_0 d\delta_0 f(\xi_0, \delta_0)\delta_D(\delta-\frac{\gamma_b}{\bar{\gamma}_b}+1)e^{i\Delta\phi(\xi_0,\delta_0)},
\end{split}
\end{equation}
where $\gamma_b$ is a function of $\xi_0$ and $\delta_0$, $N(\delta)=\int d\xi_0 d\delta_0 f(\xi_0, \delta_0)\delta_D(\delta-\frac{\gamma_b}{\bar{\gamma}_b}+1)$. Again, $I_2$ has the same form as $I_1$ with $\Delta\phi$ replaced by $2\Delta\phi$. Growth in the energy slice emittance is due to particle exchange between energy slices as a consequence of imperfect loading of the wake.

Energy slice emittance is important for two reasons: First, projected emittance growth can be reversed down to the energy slice emittance using a suitable apochromatic beam line. An example of this technique is presented in Ref.~\cite{Lindstrom2016}. Second, the energy slice emittance influences the results of some emittance measurements, such as those taken using the butterfly technique. Third, the performance of a transverse gradient undulator depends on the energy slice emittance \cite{Huang2012,Smith1979,Baxevanis2014}.

\section{\label{sec:loading}Emittance Growth of a Beam with Imperfect Wake Loading}

In the previous sections, we develop general expressions for the projected and slice emittance growth. In this section, we work out analytic solutions for the projected and slice emittance for two situations where the witness beam does not perfectly load the wake. 

\begin{figure}[bt]
  \centering
  \includegraphics[width=3.37in]{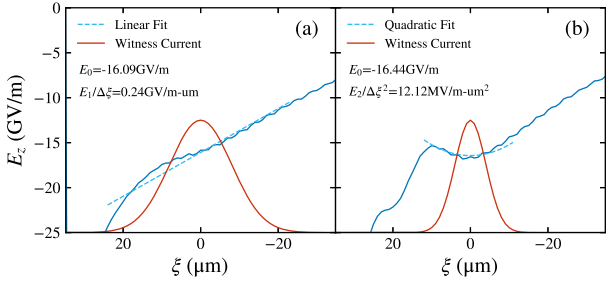}
  \caption{Longitudinal variation in the accelerating field for two witness beams of different lengths. Both beams have the same charge. The beam in (a) is too long and does not sufficiently load the wake, the accelerating field varies approximately linearly along the beam as shown by the dashed blue line. (b) shows a beam that is slightly too short, the variation in the accelerating field is well approximated by a quadratic function. In both cases the field from $-3\sigma_{\xi0}$ to $3\sigma_{\xi0}$ was used for fitting. \label{fig:loadingSim}}
\end{figure}

Perfect loading requires the witness beam to have a trapezoidal current profile to produce a uniform accelerating field \cite{Tzoufras2008}. Existing accelerators and PWFA experiments tend to use beams that have Gaussian or other non-trapezoidal current profiles. Depending on the length and current of the witness beam, the accelerating field can be approximated around the beam centroid as a linear or quadratic function of $\xi$. A linear function is a good approximation when the witness beam does not have sufficient current to flatten the wake and a quadratic function is a good approximation if the witness beam has too much current. To demonstrate this, we ran a series of particle in cell (PIC) simulations using the code VSim \cite{nieter:2004jcp}. A $0.5\,\mathrm{nC}$ witness beam and $2\,\mathrm{nC}$ drive beam were propagated through a $3.5\times10^{16}\,\mathrm{cm^{-3}}$ plasma. Both beams had Gaussian current profiles. Simulations were run for witness beams of different lengths, and thus, different peak currents. Fig.~\ref{fig:loadingSim} shows the longitudinal electric field in the region of the witness beam for two beams with the same charge, but bunch lengths of $\sigma_z=4.0\,\mathrm{\mu m}$ and $\sigma_z=8.0\,\mathrm{\mu m}$. The electric field around the shorter beam is well represented by a quadratic function centered on the beam centroid while the field around the longer beam can be approximated as linear. We proceed to work out the projected emittance, longitudinal slice emittance, and energy slice emittance growth for both cases.

In the linear case, the accelerating field is given by $E_z=-E_0-E_1\xi/\Delta\xi$. Here, $E_0$ is the accelerating field at $\xi=0$ and $E_1/\Delta\xi$ describes the slope of the electric field; $\Delta\xi$ is an arbitrary length scale, we typically set it to the bunch length. The energy of a particle in the beam as a function of $s$ depends on $\xi_0$ and $\delta_0$ according to
\begin{equation} \label{eq:gamma_linear}
\gamma_b = \bar{\gamma}_b+\delta_0\gamma_{b0}+E_1s\frac{\xi_0}{\Delta\xi}\frac{e}{m_ec^2},
\end{equation}
where $\bar{\gamma}_b=\gamma_{b0}+E_0se/(m_ec^2)$. It is straightforward to evaluate Eq.~(\ref{eqn:phi}) to get the betatron phase advance of each particle
\begin{equation*}
\phi = \frac{\omega_p}{c\sqrt{2}}\frac{2m_e c^2}{E_0e}\frac{1}{1+E_1\xi_0/(E_0\Delta\xi)}\left(\sqrt{\gamma_b}-\sqrt{\gamma_{b0}(1+\delta_0)}\right).
\end{equation*}
If we make the reasonable assumption that $E_1\xi_0/(E_0\Delta\xi)\ll1$, then $\phi$ can be expanded in the aforementioned quantity giving
\begin{equation*}
\phi = \frac{\omega_p}{c\sqrt{2}}\frac{2m_e c^2}{E_0e}\left(1-\frac{E_1\xi_0}{E_0\Delta\xi}\right)\left(\sqrt{\gamma_b}-\sqrt{\gamma_{b0}(1+\delta_0)}\right).
\end{equation*}
Even for simple distributions, the $C$ and $S$ integrals are not closed form without further expansion of $\phi$. We use the fact that $\gamma_b\approx\bar{\gamma}_b$ to expand the square root in the above expression in $E_1\xi_0/(E_0\Delta\xi)$ and $\delta_0$:
\begin{equation} \label{eq:phi_linEz}
\phi = \bar{\phi}-\delta_0\frac{\bar{\phi}}{2}\sqrt{\frac{\gamma_{b0}}{\bar{\gamma}_b}}+\frac{E_1 \xi_0}{E_0\Delta\xi}\left(\bar{k}_\beta s-\bar{\phi}\right),
\end{equation}
where we have dropped terms higher than first order. The betatron phase advance of the reference particle is 
\begin{equation} \label{eqn:barphi_linear}
\bar{\phi} = \frac{\omega_p}{c\sqrt{2}}\frac{2m_e c^2}{E_0e}\left(\sqrt{\bar{\gamma}_b}-\sqrt{\gamma_{b0}}\right).
\end{equation}
Because $\phi$ depends linearly on $\xi_0$ and $\delta_0$ as $\Delta\phi=-\delta_0\omega_\delta-\xi_0\omega_\xi$, the $I$ integrals are related to the 2D Fourier transform of the distribution:
\begin{equation} \label{eqn:CSLinear}
I_1 = \hat{f}(\omega_\xi, \omega_\delta),\quad
I_2 = \hat{f}(2\omega_\xi, 2\omega_\delta),
\end{equation}
where 
\begin{equation} \label{eqn:omegaLinear}
\begin{aligned}
&\omega_\xi=-\frac{E_1}{E_0\Delta\xi}(\bar{k}_\beta s-\bar{\phi}), \\
&\omega_\delta=\frac{\bar{\phi}}{2}\sqrt{\frac{\gamma_{b0}}{\bar{\gamma}_b}}.
\end{aligned}
\end{equation}

Consider a beam with a Gaussian longitudinal distribution with length $\sigma_{\xi0}$ and energy spread $\sigma_{\delta0}$,
\begin{equation} \label{eqn:bi-gaussian}
f(\xi_0, \delta_0) = \frac{1}{2\pi\sigma_{\xi0}\sigma_{\delta0}}\exp\left(-\frac{\xi_0^2}{2\sigma_{\xi0}^2}-\frac{\delta_0^2}{2\sigma_{\delta0}^2}\right).
\end{equation}
The distribution is symmetric so $\hat{f}$ is real and $\psi_1=\psi_2=0$. $H_1$ and $H_2$ are given by
\begin{equation}
\begin{split}
H_1&=e^{-\sigma_{\xi0}^2\omega_\xi^2/2-\sigma_{\delta 0}^2\omega_\delta^2/2}, \\
H_2&=e^{-2\sigma_{\xi0}^2\omega_\xi^2-2\sigma_{\delta 0}^2\omega_\delta^2},
\end{split}
\end{equation}
and the projected emittance and moment evolution is given by Eq.~(\ref{eq:emGrowth}) and Eqs.~(\ref{eqn:moments})-(\ref{eqn:moments_end}), respectively.

\begin{figure}[bt]
  \centering
  \includegraphics[width=3.37in]{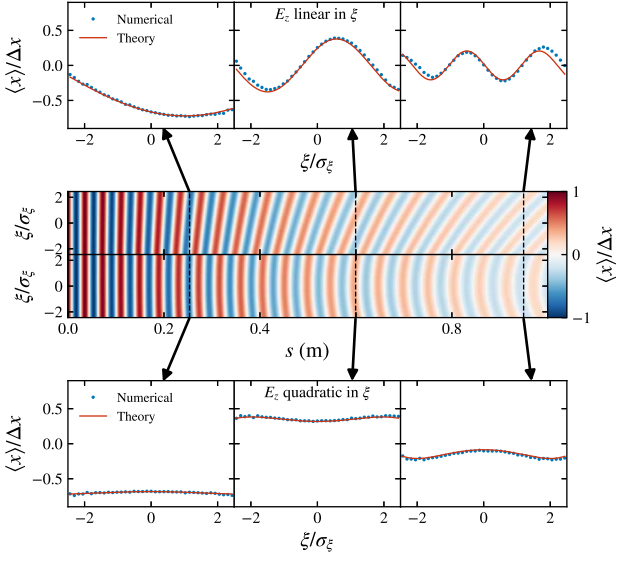}
  \caption{Centroid oscillations for each longitudinal slice in the beam. The top of the figure shows the evolution for the longer beam shown in Fig.~\ref{fig:loadingSim}(a), and the bottom shows the shorter beam from Fig.~\ref{fig:loadingSim}(b). In both cases the initial energy spread is $\sigma_{\delta0}=4\%$. The magnitude of the oscillations is dampened by chromatic phase spreading due to the initial energy spread within each slice. The oscillations decohere due to the energy differences the slices pick up as a result of imperfect wake loading. In (a) the accelerating field varies significantly across the slices and the oscillations rapidly decohere. In (b) the accelerating field along the beam is more uniform and the beam oscillates nearly coherently.  \label{fig:longSlice}}
\end{figure}

The longitudinal slice emittance of a Gaussian beam is straightforward to calculate. Evaluating Eq.~(\ref{eqn:longslice}) gives $H_1=e^{-\sigma_{\delta 0}^2\omega_\delta^2/2}$ and $\psi_1=-\omega_\xi\xi$. $H_2$ and $\psi_2$ are given by replacing $\omega_\delta$ with $2\omega_\delta$ and $\omega_\xi$ with $2\omega_\xi$ and the emittance growth is given by Eq.~(\ref{eq:emGrowth}). The longitudinal slice emittance has no dependence on $\xi$ because all the slices have the same initial energy spread. The phase term $\psi_1$ shows that the beam offset will vary sinusoidally along the bunch as shown in the top of Fig.~\ref{fig:longSlice} because $\psi_1\propto\xi_0$. The oscillation frequency is given by $\omega_\xi$.

When finding the energy slice emittance, it will be convenient to use the RMS energy spread:
\begin{equation}
\sigma_\delta^2=\sigma_{\delta 0}^2\frac{\gamma_{b0}^2}{\bar{\gamma}_b^2}+\sigma_{\xi0}^2\frac{s^2E_1^2e^2}{\bar{\gamma}_b^2\Delta\xi^2m_e^2c^4}.
\end{equation}
Using Eq.~(\ref{eqn:deltaslice}), we calculate $H_1$ and $\psi_1$ for each energy slice:
\begin{equation}
\begin{aligned}
&H_1=\exp\left[-\frac{\sigma_{\xi0}^2\sigma_{\delta 0}^2}{2\sigma_\delta^2}\left(\frac{sE_1e}{\bar{\gamma}_b\Delta\xi m_ec^2}\omega_\delta-\frac{\gamma_{b0}}{\bar{\gamma}_b}\omega_\xi\right)^2\right] \\
&\psi_1=\delta\frac{1}{\sigma_\delta^2}\left(\frac{\gamma_{b0}}{\bar{\gamma}_b}\sigma_{\delta 0}^2\omega_\delta+\frac{sE_1e}{\bar{\gamma}_b\Delta\xi m_ec^2}\sigma_{\xi0}^2\omega_\xi\right).
\end{aligned}
\end{equation}
$H_2$ and $\psi_2$ are given by replacing $\omega_\delta$ with $2\omega_\delta$ and $\omega_\xi$ with $2\omega_\xi$. The emittance growth is given by Eq.~(\ref{eq:emGrowth}) and the moments are given by Eqs.~(\ref{eqn:moments})-(\ref{eqn:moments_end}).

As with the longitudinal slice emittance, the energy slice emittance is the same for all slices because $H_1$ is independent of $\delta$ and the betatron phase offset is linearly proportional to the slice energy because $\psi_1$ is proportional to $\delta$. Chromatic dephasing cannot occur within an energy slice unless particles are able to mix between slices. The variation in accelerating field along the wake causes this mixing to occur.

There are several emittance saturation length scales. The transverse offset and mismatch driven emittance growth are determined by $H_1$ and $H_2$ respectively. We define the saturation length as the distance $s$ when the argument of the exponential in $H$ equals -1. The saturation length for the longitudinal slice emittance growth due to mismatch is given by $2\sigma_{\delta 0}^2\omega_\delta^2=1$; this length also quantifies the contribution of the initial energy spread to the emittance growth. The length scale for the emittance growth due to imperfect wake loading is $2\sigma_{\xi0}^2\omega_\xi^2=1$. Finally, the saturation length for the energy slice emittance growth is given by
\begin{equation*}
\frac{2\sigma_{\xi0}^2\sigma_{\delta 0}^2}{\sigma_\delta^2}\left(\frac{sE_1e}{\bar{\gamma}_b\Delta\xi m_ec^2}\omega_\delta-\frac{\gamma_{b0}}{\bar{\gamma}_b}\omega_\xi\right)^2=1.
\end{equation*}
The solution for $s$ for each of these expressions is summarized in Table~\ref{table:sat}. To calculate the saturation length for the energy slice emittance, we assumed $\sigma_{\delta0}^2\gamma_{b0}^2\ll\sigma_{\xi0}^2s^2E_1^2e^2/(\Delta\xi^2m_e^2c^4)$ (i.e. the initial energy spread is smaller than the energy spread induced by the wake), to simplify the expression for $\sigma_\delta$. This assumption is typically well satisfied for a plasma based accelerator with a single stage of reasonable length.

\begin{table*}[hbt]
\centering
\caption{Saturation lengths for the projected and slice emittances. The projected emittance growth is driven by contributions from both the initial energy spread and the energy spread induced by imperfect wake loading. The shorter of the two saturation lengths dominates the projected emittance growth. In contrast, both energy and longitudinal slice emittance growth depend only on the initial energy spread, assuming the variation in accelerating field is sufficiently strong.}
\label{table:sat}
 \begin{tabular}[c]{m{0.3\linewidth}m{0.35\linewidth}m{0.35\linewidth}} 
  
 \hline\hline
 \multicolumn{3}{c}{Accelerating field linear in $\xi$} \\
  & Saturation Length - Offset Beam & Saturation Length - Mismatched Beam \\
 \hline
  Imperfect Loading Contribution$^a$ & \(\displaystyle
     s_{1\xi}=\frac{E_0e}{\gamma_{b0}m_ec^2}L+\sqrt{\left(\frac{E_0e}{\gamma_{b0}m_ec^2}L\right)^2+2L}
 \) &
 \(\displaystyle
     s_{2\xi}=\frac{E_0e}{2\gamma_{b0}m_ec^2}L+\sqrt{\left(\frac{E_0e}{2\gamma_{b0}m_ec^2}L\right)^2+L}
 \) \\[5ex]

 \makecell[l]{Longitudinal Slice Emittance \\ (Initial Energy Spread Contribution)} &  \(\displaystyle
     s_{1\delta0}=4m_ec^2\gamma_{b0}\frac{m_ec\sigma_{\delta0}\omega_p\sqrt{\gamma_{b0}}-eE_0}{\left(m_ec\sigma_{\delta0}\omega_p\sqrt{\gamma_{b0}}-2eE_0\right)^2}
 \) &
 \(\displaystyle
     s_{2\delta0}=4m_ec^2\gamma_{b0}\frac{2m_ec\sigma_{\delta0}\omega_p\sqrt{\gamma_{b0}}-eE_0}{\left(2m_ec\sigma_{\delta0}\omega_p\sqrt{\gamma_{b0}}-2eE_0\right)^2}
 \) \\[5ex]

 Energy Slice Emittance$^b$ & \(\displaystyle
     s_{1\delta}\approx8m_ec^2\gamma_{b0}\frac{m_ec\sigma_{\delta0}\omega_p\sqrt{\gamma_{b0}}}{\left(m_ec\sigma_{\delta0}\omega_p\sqrt{\gamma_{b0}}-2eE_0\right)^2}
 \) &
 \(\displaystyle
     s_{2\delta}\approx8m_ec^2\gamma_{b0}\frac{2m_ec\sigma_{\delta0}\omega_p\sqrt{\gamma_{b0}}}{\left(2m_ec\sigma_{\delta0}\omega_p\sqrt{\gamma_{b0}}-2eE_0\right)^2}
 \) \\[5ex]
\hline
\multicolumn{3}{c}{Accelerating field quadratic in $\xi$} \\
\hline
Imperfect Loading Contribution$^c$ & \(\displaystyle
     s_{1\xi}=\frac{E_0e}{\gamma_{b0}m_ec^2}Q+\sqrt{\left(\frac{E_0e}{\gamma_{b0}m_ec^2}Q\right)^2+2Q}
 \) &
 \(\displaystyle
     s_{2\xi}=\frac{E_0e}{2\gamma_{b0}m_ec^2}Q+\sqrt{\left(\frac{E_0e}{2\gamma_{b0}m_ec^2}Q\right)^2+Q}
 \) \\[5ex]

 \makecell[l]{Longitudinal Slice Emittance \\ (Initial Energy Spread Contribution)} & \(\displaystyle
     s_{1\delta0}=4m_ec^2\gamma_{b0}\frac{m_ec\sigma_{\delta0}\omega_p\sqrt{\gamma_{b0}}-eE_0}{\left(m_ec\sigma_{\delta0}\omega_p\sqrt{\gamma_{b0}}-2eE_0\right)^2}
 \) &
 \(\displaystyle
     s_{2\delta0}=4m_ec^2\gamma_{b0}\frac{2m_ec\sigma_{\delta0}\omega_p\sqrt{\gamma_{b0}}-eE_0}{\left(2m_ec\sigma_{\delta0}\omega_p\sqrt{\gamma_{b0}}-2eE_0\right)^2}
 \) \\[5ex]
 \hline\hline
\end{tabular}
\begin{flushleft}
\footnotesize{$^a$ $L=\frac{E_0^2\Delta\xi^2}{E_1^2\sigma_{\xi0}^2k_{\beta0}^2}+2\sqrt{2}\frac{m_ec^2\gamma_{b0}\Delta\xi}{k_{\beta0}E_1e\sigma_{\xi0}}$. $^b$ Assuming $E_1$ is sufficiently large $\sigma_{\delta0}^2\gamma_{b0}^2\ll\sigma_{\xi0}^2s^2E_1^2e^2/(\Delta\xi^2m_e^2c^4)$.\\ $^c$ $Q=4\frac{E_0^2\Delta\xi^4}{E_2^2\sigma_{\xi0}^4k_{\beta0}^2}+8\frac{m_ec^2\gamma_{b0}\Delta\xi^2}{k_{\beta0}E_2e\sigma_{\xi0}^2}$}\\
\end{flushleft}
\end{table*}

The saturation lengths reveal several properties about the emittance growth. First, the energy slice emittance always has a saturation length longer than that of the longitudinal slice emittance. Consequently, the energy slice emittance is smaller than the longitudinal slice emittance at every $s$. Second, both slice emittances are approximately independent of $\sigma_{\xi0}$ and $E_1$; they depend only on the initial energy spread $\sigma_{\delta0}$. Third, if the initial energy spread is too low, the slice emittances will never fully saturate because acceleration reduces the relative energy spread faster than dephasing can occur. Mathematically, this appears as a divergence in the saturation lengths of the slice emittances. Saturation occurs because the spread in betatron phase becomes larger than $2\pi$; the phase spread in a longitudinal slice is $\Delta\phi=\sigma_{\delta0}\omega_\delta$, which does not grow without bound. The maximum phase spread is found by taking the large $s$ limit:
\begin{equation}
\lim_{s\to\infty}\sigma_{\delta0}\omega_\delta(s)=\sigma_{\delta0}\gamma_{b0}k_{\beta0}\frac{m_ec^2}{E_0e}.
\end{equation}
The maximum emittance is found by inserting this limit into the expression for $H_1$ and $H_2$. This also means the magnitude of the transverse oscillations of the longitudinal slices are not dampened to zero. The above conclusions apply to the energy slice emittance as long as $E_1$ is sufficiently large ($\sigma_{\delta0}^2\gamma_{b0}^2\ll\sigma_{\xi0}^2s^2E_1^2e^2/(\Delta\xi^2m_e^2c^4)$).

\begin{figure}[bt]
  \centering
  \includegraphics[width=3.37in]{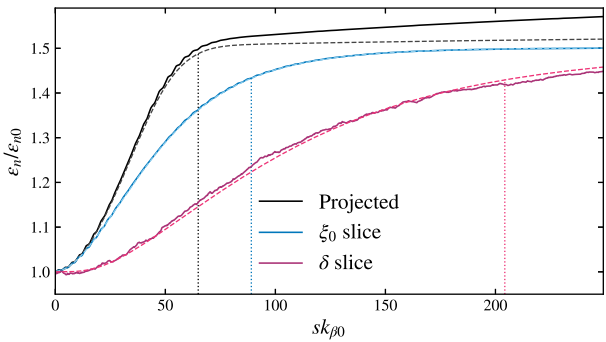}
  \caption{Growth of the projected, longitudinal slice, and energy slice emittance of the witness beam in a wake where the accelerating field varies linearly in $\xi$. The solid lines are the emittance from numerical particle tracking, the dashed lines are the analytic theory. The dotted lines show the saturation lengths from Table~\ref{table:sat}: black is the contribution from the non-uniform accelerating field to the projected emittance, blue is the initial energy spread contribution as well as the longitudinal slice, and magenta is the energy slice. The projected emittance is dominated by contribution from the non-uniform accelerating field. \label{fig:sliceEmt}}
\end{figure}

Figure~\ref{fig:sliceEmt} compares the growth of the longitudinal slice, energy slice, and projected emittance. The accelerating field is the same as that shown in Fig.~\ref{fig:loadingSim}(a), but the beam length has been doubled to $\sigma_{\xi0}=16.0\,\mathrm{\mu m}$ to exaggerate the difference between the projected and slice emittance saturation lengths. The initial energy spread is $\sigma_{\delta0}=0.05$. The saturation length due to the variation in the accelerating field is shorter than that due to the beam's initial energy spread. As a result, the saturation length due to imperfect wake loading adequately describes the projected emittance growth. The difference between the numerical and analytic solution for the projected emittance arises due to the large final energy spread in this example. This growth can be handled analytically using the approach presented in Ref.~\cite{ariniello:2019prab}; however, for typical experimental parameters the energy spread is small enough that the correction is negligible. 

For the quadratic case, the accelerating field is given by $E_z=-E_0-E_2\xi^2/\Delta\xi^2$, where $E_2/\Delta\xi^2$ is the quadratic fitting parameter. As before, $\Delta\xi$ is an arbitrary length scale typically set to the bunch length. The energy of a particle in the beam is 
\begin{equation}
\gamma_b=\bar{\gamma}_b+\delta_0\gamma_{b0}+E_2 s \frac{\xi_0^2}{\Delta\xi^2}\frac{e}{m_ec^2},
\end{equation}
where $\bar{\gamma}_b\gamma_{b0}+E_0se/(m_ec^2)$, the same as before. Following the same procedure as before, the betatron phase advance of the particle is given by
\begin{equation}
\phi=\bar{\phi}-\delta_0\frac{\bar{\phi}}{2}\sqrt{\frac{\gamma_{b0}}{\bar{\gamma}_b}}+\frac{E_2\xi_0^2}{E_0\Delta\xi^2}(\bar{k}_\beta s-\bar{\phi}).
\end{equation}
The reference phase advance $\bar{\phi}$ is given by Eq.~(\ref{eqn:barphi_linear}). Unlike before, $\phi$ does not depend linearly on $\xi_0$ and thus the $C$ and $S$ integrals are no longer Fourier transforms. To keep the math simple, we define the $\mathrm{w}_\xi$ as
\begin{equation}
    \mathrm{w}_\xi=-\frac{E_2}{E_0\Delta\xi^2}(\bar{k}_\beta s-\bar{\phi})
\end{equation}
while $\omega_\delta$ is given by Eq.~(\ref{eqn:omegaLinear}).

We again assume the Gaussian longitudinal phase space distribution of Eq.~(\ref{eqn:bi-gaussian}). In this case, the $I$ integrals are straightforward and the evolution is described by
\begin{equation}
\begin{aligned}
&H_1=\left(1+4\sigma_{\xi0}^4\mathrm{w}_\xi^2\right)^{-1/4}e^{-\sigma_{\delta0}^2\omega_\delta^2/2} \\
&\psi_1(\mathrm{w}_\xi, \omega_\delta)=\frac{1}{2}\mathrm{arctan}\left(2\sigma_{\xi0}^2\mathrm{w}_\xi\right).
\end{aligned}
\end{equation}
$H_2$ and $\psi_2$ are given by making the replacement $\omega_\delta\rightarrow2\omega_\delta$ and $\mathrm{w}_\xi\rightarrow2\mathrm{w}_\xi$. In this case $\psi$ does not strictly satisfy the requirements for Eq.~(\ref{eq:emGrowth}) to be valid; however, $2\sigma_{\xi0}^2\mathrm{w}_\xi$ must be small when the emittance is still growing (the emittance saturation length for mismatch is given by $\sigma_{\xi0}^2\mathrm{w}_\xi=2$), allowing us to expand the arctan and satisfy the condition that $\psi_2=2\psi_1$. To calculate the saturation lengths, we let $\left(1+16\sigma_{\xi0}^4\mathrm{w}_\xi^2\right)^{-1/4}\approx e^{-1}$ for mismatch ($H_2)$ which corresponds to $\sigma_{\xi0}^2\mathrm{w}_\xi=2$. For transverse offset we assume $\sigma_{\xi0}^2\mathrm{w}_\xi=4$ in $H_1$ in order to calculate the saturation length. The saturation lengths are shown in Table~\ref{table:sat}.

As before, all longitudinal slices have the same slice emittance. The evolution of each longitudinal slice is given by $H(\omega_\delta)=e^{-\sigma_{\delta 0}^2\omega_\delta^2/2}$ and $\psi(\mathrm{w}_\xi)=-\mathrm{w}_\xi\xi^2$. The saturation length is the same as in the linear loading case. If the bunch starts offset, the $\xi^2$ dependence of the phase leads to a large region in the center of the bunch that undergoes transverse oscillations in phase. This is evident in the bottom of Fig.~\ref{fig:longSlice}. 

\begin{figure}[bt]
  \centering
  \includegraphics[width=3.37in]{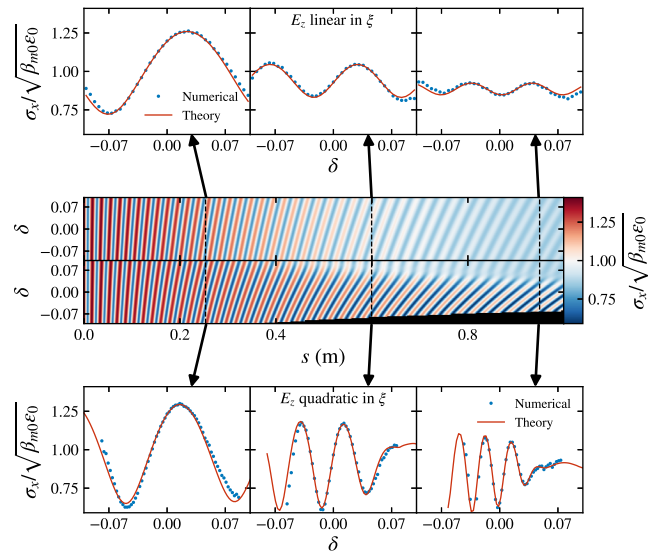}
  \caption{Spot size for each energy slice in the beam. The top of the figure shows the evolution for the longer beam shown in Fig.~\ref{fig:loadingSim}(a) and the bottom shows the shorter beam from Fig.~\ref{fig:loadingSim}(b). In both cases the initial energy spread is $\sigma_{\delta0}=4\%$. Chromatic dephasing cannot occur within a slice; the dampening of the $\beta$ function oscillations is due to particle exchange between slices driven by wake loading. If $E_z$ is linear in $\xi$ (top), the emittance grows uniformly for all slices and the beam size only varies with the $\beta$ function of each slice. If $E_z$ is quadratic in $\xi$ (bottom), the emittance varies across the slices leading to more complex dynamics. \label{fig:deltaSlice}}
\end{figure}

The energy slice emittance is analytically tractable, but the solution is cumbersome and the $H$ and $\psi$ functions are not easily extracted. To compare to the linear case, we use numerical particle tracking to propagate a witness beam in the accelerating field from Fig.~\ref{fig:loadingSim}(b) and then numerically evaluate Eq.~(\ref{eqn:deltaslice}) to get the theoretical prediction. For this example, the beam is mismatched but not transversely offset. Fig.~\ref{fig:deltaSlice} shows a comparison of the energy slice spot size evolution between the linear and quadratic cases. In both cases the spot size varies across the energy slices at the exit of the plasma. Unlike the linear loading case, the emittance in the quadratic loading case varies across the energy slices. This variation is described by the $H_2$ function which is visible in the figure as a $\delta$ dependent dampening of spot size oscillation. The low energy slices of the beam disappear because particles can only move to slices with larger $\delta$ (if the initial energy spread is uncorrelated). 

The variation in beam parameters with energy could be used to measure the wake loading. If the beam is intentionally mismatched into the plasma, the C-S parameters and emittance of each energy slice will have a dependence on the wake loading. An imaging spectrometer can then indirectly measure the loading by looking at the variation in $\sigma_x$ with energy. 

\section{Emittance Evolution in a Plasma Source with Density Ramps}

In the previous examples, we have only considered plasma sources with uniform density. Here, we analytically calculate the quantities necessary to find the emittance growth in a plasma source with adiabatic density ramps at the entrance and exit. We assume the accelerating field varies with plasma density according to the simple model
\begin{equation}
E = E_0\left(\sqrt{\eta}-2\eta\right)-E_1\eta\frac{\xi_0}{\Delta\xi},
\end{equation}
where $\eta=n_e/n_{eu}$, $n_{eu}$ is the uniform density in the bulk, $E_1\eta/\Delta\xi$ is the slope of the wakefield in the blowout regime \cite{Lu2006, Lu2006a} and $E_0\left(\sqrt{\eta}-2\eta\right)$ describes the variation in longitudinal phase and maximum wake amplitude with plasma density \cite{Litos2019}. The fields can be approximated from first principles as 
\begin{equation} \label{eq:E0E1}
E_0=\frac{\kappa}{2}\pi c\sqrt{\frac{n_{eu}m_e}{\epsilon_0}},\quad \frac{E_1}{\Delta\xi}= \frac{n_{eu}e}{2\epsilon_0},
\end{equation}
where $\kappa$ is a multiplier accounting for the current of the drive beam. $\kappa$ typically varies between 1 and 2. Beam loading is ignored in this example to keep the math tractable.

The energy of a single particle is then given by
\begin{equation}
\begin{split}
\gamma_b = &\gamma_{b0}+\delta_0\gamma_{b0}-\frac{E_0e}{m_ec^2}G_1 \\
&+\frac{e}{m_ec^2}\left(2E_0+E_1\frac{\xi_0}{\Delta\xi}\right)G_2,
\end{split}
\end{equation}
where
\begin{equation}
G_1(s)=\int_{0}^sds'\sqrt{\eta(s')},\quad G_2(s)=\int_{0}^sds'\eta(s').
\end{equation}
To find the betatron phase advance in the ramp we assume the energy spread and energy gained in the ramp is small compared to the beam's centroid energy throughout the ramp. We can then expand $k_\beta$ as
\begin{equation*}
k_\beta=\frac{\omega_{pu}}{c\sqrt{2\gamma_{b0}}}\eta\left(1-\frac{1}{2}\frac{\Delta\gamma_b}{\gamma_{b0}}\right),
\end{equation*}
where $\Delta\gamma_b=\gamma_b-\gamma_{b0}$ and $\omega_{pu}^2=n_{eu}e^2/m_e\epsilon_0$ is the plasma frequency of the uniform density region. As before, the betatron phase advance is linear in $\delta_0$ and $\xi_0$: $\Delta\phi=-\delta_0\omega_\delta-\xi_0\omega_\xi$. The $I$ integrals are given by Eq.~(\ref{eqn:CSLinear}) with
\begin{equation}
\begin{aligned}
&\omega_\xi=\frac{\omega_{pu}}{c\sqrt{2\gamma_{b0}}}\frac{E_1e}{2m_ec^2\gamma_{b0}\Delta\xi}D_2 \\
&\omega_\delta=\frac{1}{2}\frac{\omega_{pu}}{c\sqrt{2\gamma_{b0}}}G_1.
\end{aligned}
\end{equation}
The betatron phase advance of the reference particle is
\begin{equation}
\bar{\phi}=\frac{\omega_{pu}}{c\sqrt{2\gamma_{b0}}}\left[G_1-\frac{E_0e}{2m_ec^2\gamma_{b0}}(2D_2-D_1)\right],
\end{equation}
where
\begin{equation}
D_1=\int_{0}^sds'\sqrt{\eta(s')}G_1(s'),\quad D_2=\int_{0}^sds'\sqrt{\eta(s')}G_2(s').
\end{equation}
The emittance growth and beam moments are then found by getting $H_1$, $H_2$, $\psi_1$, and $\psi_2$ from the $I$ integrals and then using Eq.~(\ref{eq:emGrowth}) to find the emittance and Eqs.~(\ref{eqn:moments})-(\ref{eqn:moments_end}) to find the moments. Without carrying out the full calculation, it is apparent that minimizing the $G$ and $D$ integrals will minimize $\omega_\xi$, $\omega_\delta$, and $\Delta\gamma_b$, i.e., minimize the undesirable impacts the ramp has on the beam.

As the beam enters the uniform density region, it is accelerated significantly and we can no longer assume $\Delta\gamma_b\ll\gamma_{b0}$. Instead, we use the solution presented in Sec.~\ref{sec:loading} with the addition of an initial phase and an initial energy of $\gamma_0\rightarrow\gamma_{bl}=\bar{\gamma}_{b}(s=l)$, where $l$ is the length of the ramp. The exit ramp is handled the same way as the entrance ramp except the initial energy is $\gamma_{bL}=\bar{\gamma}_{b}(s=l+L)$, where $L$ is the length of the uniform plasma. The resulting piece-wise functions for $\bar{\phi}$, $\omega_\xi$, and $\omega_\delta$ are fairly cumbersome and are given in Appendix~\ref{app:loading}. 

\begin{figure}[bt]
  \centering
  \includegraphics[width=3.37in]{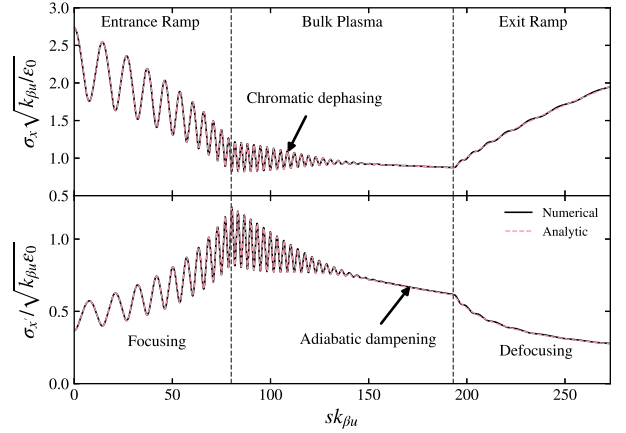}
  \caption{Evolution of the spot size (top) and divergence (bottom) of a mismatched beam entering a plasma source with entrance and exit ramps. The beam is initially focused by the entrance ramp into the bulk of the plasma source where it undergoes chromatic dephasing and becomes matched to the plasma. As the beam is accelerated, the divergence continues to decrease due to adiabatic dampening before the beam is defocused by the exit ramp. \label{fig:PlasmaWithRamps}}
\end{figure}

Qualitatively, the solution is similar to that of a uniform plasma presented in Sec.~\ref{sec:loading}. The primary difference is the conversion of beam size into divergence by the entrance ramp (focusing) and divergence into beam size by the exit ramp (defocusing). This is shown in Fig.~\ref{fig:PlasmaWithRamps} where $\sigma_x$ and $\sigma_{x'}$ are plotted for a mismatched beam propagating through a plasma source with ramps. In this example, the ramps are short and only a small amount of chromatic dephasing occurs in them. The majority of the emittance growth occurs in the bulk plasma.

\begin{figure}[bt]
  \centering
  \includegraphics[width=3.37in]{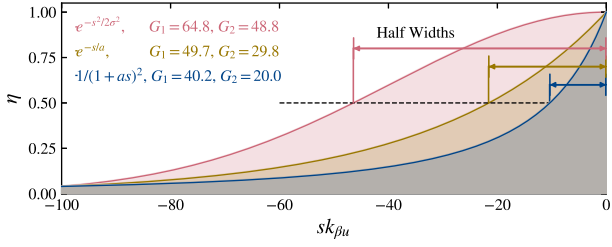}
  \caption{Three plasma ramps with different profiles that all reduce the beta function by a factor of 5. The chromatic phase spread, betatron oscillations and drive beam energy loss in the ramp all depend on the integral over the plasma density ($G_2$ shaded area). The first ramp (blue, bottom) is the most efficient adiabatic ramp possible while the Gaussian (red, top) is barely adiabatic at the tail. The ramp shape can have a significant impact on the beam, and the ramp half width is not a good indicator of the effectiveness of the ramp. \label{fig:RampShapes}}
\end{figure}

If the ramps are long or poorly shaped, they can have significant impacts on the witness beam. The integrals $G_1$, $G_2$, $D_1$, and $D_2$ determine the amount of phase spread, betatron oscillation, and, in the case of a beam driven wake, drive beam energy loss in the ramp. All the integrals are reduced if the integrated plasma density---$G_2$---is minimized. The adiabatic ramp that minimizes the impact on the beam therefore has the highest density gradient possible while remaining adiabatic. Solving Eq.~(\ref{eq:Aparam}) gives the optimal ramp as $\eta=1/(1-2\alpha_ms)^2$. From experience, the ramp needs $\abs{\alpha_m}<0.1$ for the ramp to be well described by adiabatic theory. In Fig.~\ref{fig:RampShapes}, we compare several different adiabatic ramp shapes that all focus the beam by the same amount. The ramp shapes that minimize the density integral induce smaller impacts on the witness beam. The full ramp shape is important and the impact of the ramp on the beam cannot be described using only the half width and the adiabatic parameter. 

\section{Particle Injection into a Linear Accelerating Field}

Consider an injector where a low charge beam is generated within the wake. The following discussion is agnostic to the details of the injection scheme; for example, it could be applied equally well to either ionization injection or plasma photo-cathode injection \cite{Pak2010,Hidding2012}. We can use our theory to extend the theoretical treatment presented by Ref.~\cite{Xu2014a} to include arbitrary distributions of injected electrons and offset of the injected beam with respect to the center of the wake. 

Assume at time $\tau_1$ a group of electrons is injected into the wake and rapidly accelerated until each electron reaches a phase locked longitudinal position $\xi_0$ with energy $\gamma_{b0}$. After the particles are phase locked, the longitudinal distribution of the particles is given by $f(\xi_0, \tau_1)$, and the transverse distribution of the particles is described by an initial set of CS parameters $\beta_0$, $\alpha_0$, and $\gamma_0$ and an initial transverse emittance $\epsilon_{n0}$. Assume particles ionized at a different time $\tau_2$ are described by the same CS parameters and emittance once they are phase locked. The longitudinal phase space of particles injected at $\tau_2$ is given by $f(\xi_0, \tau_2)$.

Further, assume a small amount of charge is injected into the wake, and thus the longitudinal electric field $E_z$ varies linearly along the length of the injected beam, $E_z=-E_0-E_1\xi/\Delta\xi$. The energy of an injected particle depends on the injection time and the phase within the wake according to
\begin{equation} \label{eq:gamma_inj}
\gamma_b = \bar{\gamma}_b-E_0c\tau\frac{e}{m_ec^2}+E_1\frac{\xi_0}{\Delta\xi}(s-c\tau)\frac{e}{m_ec^2},
\end{equation}
where $\bar{\gamma}_b=\gamma_{b0}+E_0se/(m_ec^2)$ and $\gamma_{b0}$ is the energy at the moment of phase locking. The wake is moving at $c$, so a particle injected at $\tau$ is phase locked into position $\xi_0$ at a location of $s_0=\tau c$ . Using the same approach and assumptions as in Sec.~\ref{sec:loading} to evaluate Eq.~(\ref{eqn:phi}) gives the betatron phase advance of each particle
\begin{equation} \label{eq:phi_inj}
\phi = \bar{\phi}-\bar{k}_\beta c\tau+\frac{E_1 \xi_0}{E_0\Delta\xi}\left(\bar{k}_\beta s-\bar{\phi}\right),
\end{equation}
where $\bar{\phi}$ is given by Eq.~(\ref{eqn:barphi_linear}).

The $C$ and $S$ integrals are now in terms of $\tau$ rather than $\delta$, we only write $C_1$ for brevity:
\begin{equation}
C_1 = \int d\xi_0d\tau f(\xi_0, \tau)\cos[\phi(\xi_0, \tau)].
\end{equation}
Because $\phi$ depends linearly on $\tau$ and $\xi$, Eq.~(\ref{eqn:CSLinear}) holds with $\delta$ replaced by $\tau$. $\omega_\xi$ remains the same as in Eq.~(\ref{eqn:omegaLinear}) while $\omega_\tau=\bar{k}_\beta c$.

As pointed out in Ref.~\cite{Xu2014a}, the projected emittance initially rises rapidly during injection before decreasing to a minimum and finally growing to saturation. We can solve for the propagation distance that minimizes the emittance growth by finding the value of $s$ where $H_1$ and $H_2$ are maximized. Further, we can solve for the propagation distance that minimizes the energy spread. In an optimal injector, the energy spread and emittance will reach their minimal values simultaneously at the exit of the accelerator. 

It is straightforward to solve for the energy spread starting with Eq.~(\ref{eq:gamma_inj}). Dropping moments of the longitudinal distribution higher than order 2 gives:
\begin{equation}
\sigma_\delta^2 = \frac{e^2}{c^2m_e^2\bar{\gamma}_b^2}\left[\frac{E_1^2s^2}{c^2\Delta\xi^2}\sigma_\xi^2+E_0^2\sigma_\tau^2-\frac{2E_0E_1s}{c\Delta\xi}\sigma_{\xi\tau}\right],
\end{equation}
where $\sigma_\xi$ and $\sigma_\tau$ are the standard deviations of the longitudinal distribution about the mean and $\sigma_{\xi\tau}=\left<\xi\tau\right>-\left<\xi\right>\left<\tau\right>$. As long as $sE_0e/m_ec^2\gg\gamma_{b0}$, the constant term $\gamma_{b0}$ in $\bar{\gamma}_b$ can be dropped, giving the explicit $s$ dependence of the relative energy spread as
\begin{equation}
\sigma_\delta^2=\frac{E_1^2}{E_0^2\Delta\xi^2}\sigma_\xi^2+\frac{c^2}{s^2}\sigma_\tau^2-\frac{2E_1c}{E_0\Delta\xi s}\sigma_{\xi\tau}.
\end{equation}
The first term is the asymptotic energy spread resulting from imperfect wake loading. The second term is the energy spread induced by the finite injection time. This energy spread is fixed; thus, its contribution to the relative energy spread is suppressed by a factor of $1/\bar{\gamma}_b$ as the beam accelerates. The final term accounts for any initial correlation between injection time and energy, particles that are injected later (earlier) experience larger (smaller) accelerating fields, thus flattening the longitudinal phase space. 

The minimum energy spread is given by 
\begin{equation}
\sigma_{\delta min}^2=\frac{E_1^2}{E_0^2\Delta\xi^2}\left[\sigma_\xi^2-\frac{\sigma_{\xi\tau}^2}{\sigma_\tau^2}\right].
\end{equation}
The beam reaches its minimum energy spread at 
\begin{equation} \label{eq:s_mindelta}
s=\frac{c\sigma_\tau^2E_0\Delta\xi}{E_1\sigma_{\xi\tau}},
\end{equation}
which requires $\sigma_{\xi\tau}>0$ in order for $s>0$, which is equivalent to an initial positive chirp. The energy spread can be minimized by creating a large correlation between injection time and injection position.

Calculating the value of the minimum emittance requires assuming an initial longitudinal distribution. Take as an example an injection process that injects particles into a blowout wake at a uniform rate from $\tau=-\Delta\tau/2$ to $\tau=\Delta\tau/2$. After becoming phase locked, the particles injected at a given $\tau$ are longitudinally distributed in the wake with a Gaussian distribution centered at $v\tau$. Here, $v$ describes the change in the longitudinal position of the particles with injection time. Positive $v$ means particles injected at later times are injected closer to the back of the wake. The distribution $f(\xi_0, \tau)$ is separable and can be written as
\begin{equation}
\begin{split}
f(\xi_0, \tau)=&\frac{1}{\sigma_{\xi0}\sqrt{2\pi}}e^{-(\xi_0-v\tau)^2/(2\sigma_{\xi0}^2)} \\
&\times\frac{1}{T}\mathrm{rect}\left[\frac{\tau}{T}-\frac{1}{2}\left(1-\frac{\Delta\tau}{T}\right)\right],
\end{split}
\end{equation}
where $T=s/c+\Delta\tau/2$ for $0\le s<c\Delta\tau/2$ and $T=\Delta\tau$ for $c\Delta\tau/2\le s$. The width $T$ accounts for the change in the distribution during injection. The instantaneous injection length $\sigma_{\xi0}$ is the bunch length of particles injected at a given $\tau$ after they become phase locked. The full bunch has a length of $\sigma_\xi^2=\sigma_{\xi0}^2T^2v^2/12$. If the injection time is sufficiently long, the energy spread of the beam can rise to to 100\% ($\sigma_\delta\approx0.5$ for our distribution); then, we can no longer assume $\beta_m$ and $\gamma_b$ are approximately equal for all particles. The minimum achievable energy spread, however, is small and our approach can be used to describe the beam around this minimum. Using Eq.~(\ref{eqn:CSLinear}) we get 
\begin{equation}
\begin{split}
H_1(\omega_\xi, \omega_\tau) &= e^{-\omega_\xi^2\sigma_{\xi0}^2/2}\mathrm{sinc}\left[\frac{T}{2}(\omega_\xi v+\omega_\tau)\right] \\
H_2(\omega_\xi, \omega_\tau) &= e^{-2\omega_\xi^2\sigma_{\xi0}^2}\mathrm{sinc}\left[T(\omega_\xi v+\omega_\tau)\right] \\
\psi_1(\omega_\xi, \omega_\tau) &= \frac{1}{2}(\Delta\tau-T)(\omega_\xi v+\omega_\tau) \\
\psi_2(\omega_\xi, \omega_\tau) &= (\Delta\tau-T)(\omega_\xi v+\omega_\tau)
\end{split}
\end{equation}
Since $2\psi_1=\psi_2$, the emittance evolution is described by Eq.~(\ref{eq:emGrowth}). 

In the case of large $v$ and small $\sigma_{\xi0}$ (strictly $1\ll Tv/\sigma_{\xi0}+T\omega_\tau/(\sigma_{\xi0}\omega_\xi)$), we can approximate the exponential in $H_1$ and $H_2$ as 1 and solve for $s$ where both $H$ functions are maximized:
\begin{equation} \label{eq:s_min_em}
s = \frac{cE_0\Delta\xi}{E_1v},
\end{equation}
which is same distance where the minimum energy spread occurs. This injection distribution works well for the accelerating field assumed here. For even moderate injection duration, however, the minimum emittance is only marginally smaller than the saturated emittance.

Using the estimates for $E_0$ and $E_1$ from Eq.~(\ref{eq:E0E1}), we can derive useful formulas for designing a plasma injector. In this case $\kappa$ can also be used to describe the wakefield phase the particles are injected into. We assume the length of the injector is chosen to minimize the final energy spread of the beam, combining Eq.~(\ref{eq:s_mindelta}) with Eq.~(\ref{eq:E0E1}) gives
\begin{equation}
L=\frac{\pi c\kappa}{\omega_p}\frac{c}{v}.
\end{equation}
The final energy spread of the injected beam is given by
\begin{equation}
\sigma_{\delta f}=\frac{\omega_p\sigma_{\xi 0}}{\pi c \kappa}
\end{equation}
and the final energy of the beam is given by
\begin{equation}
\gamma_{bf}=\gamma_{b0}+\frac{\pi^2\kappa^2}{2}\frac{c}{v}.
\end{equation}
The final energy at the point of minimum energy spread is independent of the plasma density and only depends on the correlation between $\tau$ and $\xi$. This occurs because the length of the plasma scales inversely to the plasma density and the correlation $v$. The final energy spread scales with $\sqrt{n_e}\sigma_{\xi0}$. Thus, the injection region needs to be reduced in proportion to the skin depth to maintain a given energy spread. Some injection schemes have a minimum attainable $\sigma_{\xi0}$ resulting in a trade-off between emittance and energy spread because space charge effects, and thus the initial emittance, are reduced as plasma density increases. This trade-off can potentially be mitigated by appropriately loading the wake to reduce the energy spread.

\begin{figure}[bt]
  \centering
  \includegraphics[width=3.37in]{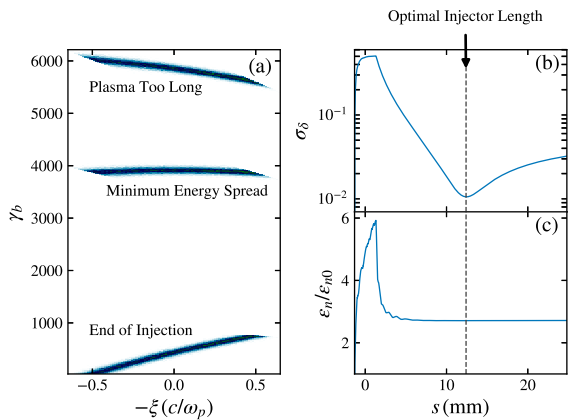}
  \caption{Evolution of the injected beam in a plasma based injector designed for 1\% energy spread at $2\,\mathrm{GeV}$. The accelerating field varies linearly along the beam. Particles are injected with a correlation between injection time and longitudinal position. (a) Immediately after injection ends, particles injected earlier have higher energy because they have experienced the accelerating field for a longer time. When the beam reaches the design energy ($2\,\mathrm{GeV}$) the energy spread is minimized as particles injected later have experienced a stronger electric field. (b) If the beam propagates too long, the energy spread starts to increase. (c) The emittance initially grows with the energy spread before settling down to just below the saturated value. \label{fig:Injection}}
\end{figure}

As an example, let us design an injector to produce $2\,\mathrm{GeV}$ beams with sub 1\% energy spread. Assume $\kappa=1.5$ and $\gamma_{b0}=25$. Immediately we can find the required correlation is $v=c/350$, this is reasonable for currently proposed injection schemes \cite{Xu2017,Dalichaouch2020}. The plasma density might depend on the target emittance or the drive beam available; we use a typical experimental value of $n_e=5\times10^{17}\,\mathrm{cm^{-3}}$. The plasma should be $12.4\,\mathrm{mm}$ long to reach minimum energy spread; to achieve sub 1\% energy spread, the particles should be injected such that $\sigma_{\xi0}\leq0.35\,\mathrm{\mu m}$; we want to emphasize that this is not the final bunch length, but the phase locked length of particles injected at a given $\tau$.  Fig.~\ref{fig:Injection} shows the evolution of the longitudinal phase space, energy spread, and projected emittance within the injector. The wake loading effectively cancels out the initial energy chirp of the beam. For parameters of interest to practical injector designs, the minimum emittance is approximately equal to the saturated emittance, thus length is not a concern for $\epsilon_n$. The initial spike in emittance is primarily due to the very large energy spread present while injection is still occurring. Notice that while the injector length is not particularly important for the final emittance, it has a significant impact on the final energy spread.

\section{\label{sec:conc}Conclusion}

We have demonstrated an analytic approach to calculating the evolution of the witness beam in a plasma based accelerator operating in the nonlinear blowout regime. We included the effects of energy change, loading of the wake, and adiabatic variations in plasma density. We developed our approach to describe the chromatic dephasing of the beam. This dephasing will cause emittance growth of a beam if the beam is either transversely offset or mismatched to the plasma. The growth will saturate if the plasma is sufficiently long. The saturated emittance is the sum of the contribution from the offset and the mismatch, with the transverse offset requiring a longer distance to saturate. In addition, we showed how to calculate both the energy slice and longitudinal slice emittance evolution and saturation values.

For simple cases, the projected and slice emittances can be calculated analytically, letting us investigate general properties of the emittances. Because the particles are phase locked in the wake, the longitudinal slice emittance depends only on the initial energy spread within the slice and grows more slowly than the projected emittance. In the presence of imperfect beam loading, the variation in accelerating field along the beam causes particles to mix between energy slices, leading to growth of the energy slice emittance. Depending on the details of the wake loading, the energy slice emittance will vary across the slices. 

In addition to the emittance, our approach provides the beam moments and thus the transverse offset and spot size of the projected beam and the longitudinal/energy slices. Chromatic dephasing leads to a dampening of any transverse offset on the same time frame as the emittance growth. For the energy slices, the mixing process results in an energy dependence of the beam spot size at the exit of the plasma. This dependence is sensitive to the details of the beam loading. This will impact the signal the electron beam generates in an imaging spectrometer, which can be used to indirectly measure the beam loading. 

We showed two examples of how our general approach can be applied to specific situations. First, we considered a full plasma accelerator with ramps but with insufficient charge in the witness beam to load the wake. In this case, the energy spread produced by the variation in the accelerating field is sufficient for the emittance to reach saturation regardless of the initial energy spread. We also showed that it is the integral of the plasma density ramp profile that determines how much the ramp perturbs the beam. Second, we considered a plasma injection scheme. For this example, we derived some simple scaling laws for the final energy spread and optimal length of the injector to simultaneously minimize the energy spread and emittance. 

In many of our examples, we considered low charge beams that do not significantly load the wake. Our approach, however, is capable of handling more complex loading situations if the longitudinal dependence of the accelerating field can be written analytically. Even if it cannot, the integrals can be solved numerically. Depending on the longitudinal phase space, this may be faster than particle tracking.

\section*{Acknowledgments}

This material is based upon work supported by the U.S. Department of Energy, Office of Science, Office of High Energy Physics under Award No. DE-SC0017906.

\appendix

\section{Particle Tracking}

The simulations shown in this paper all use a simple particle tracking code. Each particle is propagated by numerically solving Eq.~(\ref{eqn:EOM}). The values of $\gamma_b(s)$ and $\gamma_b'(s)$ are known from the acceleration model used for each simulation. The focusing force, parameterized by $k_\beta(s)$, is easily found from $\gamma_b(s)$ and the plasma density profile $n_e(s)$. Both $k_\beta(s)$ and $\gamma_b(s)$ are functions of the particles initial $\xi$, and either $\delta$ or $\tau$. The particle transverse positions are updated from step $s_i$ to $s_{i+1}$ using the transport matrix formalism
\begin{equation}
\begin{pmatrix} x_{i+1}\\ x'_{i+1} \end{pmatrix} = M(s_{i+1}|s_i) \begin{pmatrix} x_i\\ x_i' \end{pmatrix},
\end{equation}
where we use the standard transport matrix
\begin{equation}
M(s_{i+1}|s_i) = \begin{pmatrix} \cos(k_\beta \Delta s) & \frac{1}{k_\beta}\sin(k_\beta \Delta s) \\ -\theta k_\beta\sin(k_\beta \Delta s) & \theta\cos(k_\beta \Delta s) \end{pmatrix}.
\end{equation}
Here, $\Delta s=s_{i+1}-s_i$ is the step size, $k_\beta=k_\beta(s_i+\Delta s/2)$ is the betatron wavenumber evaluated at the half-step, and $\theta=1-\frac{\gamma_b'(s_i+\Delta s/2)\Delta s}{\gamma_b(s_i+\Delta s/2)}$ describes the adiabatic dampening. The step size $\Delta s$ is much less than $1/k_\beta$.

The particles are initialized in action-angle variable space ($J$ and $\psi$) using the distribution
\begingroup
\begin{align}
\rho(J) &= \frac{1}{\epsilon}e^{-J/\epsilon}, \\
\rho(\psi) &= \frac{1}{2\pi}\mathrm{rect}\left(\frac{\psi}{2\pi}\right).
\end{align}
\endgroup
The particle's initial position in real space is calculated from $J$ and $\psi$ using
\begingroup
\begin{align}
x_0 &= \sqrt{2J\beta_0}\cos\psi+\Delta x, \\
x_0' &= -\sqrt{\frac{2J}{\beta_0}}\left(\sin\psi+\alpha_0\cos\psi\right).
\end{align}
\endgroup
The longitudinal positions are initialized based on the distribution of interest.

\section{\label{app:loading} Beam Evolution in a Plasma Source with Adiabatic Ramps}

Finding the beam evolution in an adiabatic plasma source with ramps requires assuming the energy change is small in the ramps but including energy change in the bulk plasma. As a result, the analytic expressions are piece-wise with different expressions for the ramps and the bulk. The evolution is fully described by $\gamma_b$, $\bar{\phi}$, $\omega_\xi$, and $\omega_\delta$. The expression for $\gamma_b$ and $\bar{\gamma}_b$ are not piecewise and are given in the text. The expressions for $\bar{\phi}$, $\omega_\xi$, and $\omega_\delta$ are
\begin{widetext}
\[
\bar{\phi}=
\begin{cases} 
      \frac{\omega_{pu}}{c\sqrt{2\gamma_{b0}}}\left[G_1-\frac{E_0e}{2m_ec^2\gamma_{b0}}(2D_2-D_1)\right] & s\leq l \\
      \bar{\phi}|_l+\frac{\omega_{pu}}{c\sqrt{2}}\frac{2m_e c^2}{E_0e}\left(\sqrt{\bar{\gamma}_b}-\sqrt{\gamma_{bl}}\right) & l< s\leq l+L \\
      \bar{\phi}|_{l+L}+\frac{\omega_{pu}}{c\sqrt{2\gamma_{bL}}}\left[G_1^*-\frac{E_0e}{2m_ec^2\gamma_{bL}}(2D_2^*-D_1^*)\right] & l+L< s
   \end{cases}
\]

\[
\omega_\xi=
\begin{cases} 
      \frac{\omega_{pu}}{c\sqrt{2\gamma_{b0}}}\frac{E_1e}{2m_ec^2\gamma_{b0}\Delta\xi}D_2 & s\leq l \\
      \omega_\xi|_l-\frac{\omega_{pu}}{c\sqrt{2}}\frac{E_1}{E_0\Delta\xi}\left[\frac{1}{\sqrt{\bar{\gamma}_b}}G_2+\frac{2m_ec^2}{E_0e}(\sqrt{\gamma_{bl}}-\sqrt{\bar{\gamma}_b})-\frac{1}{\sqrt{\gamma_{bl}}}G_2(l)\right] & l< s\leq l+L \\
      \omega_\xi|_{l+L}+\frac{\omega_{pu}}{c\sqrt{2\gamma_{bL}}}\frac{E_1e}{2m_ec^2\gamma_{bL}\Delta\xi}\left[D_2^*+G_1^*G_2(L)\right] & l+L< s
   \end{cases}
\]

\[
\omega_\delta=
\begin{cases} 
      \frac{1}{2}\frac{\omega_{pu}}{c\sqrt{2\gamma_{b0}}}G_1 & s\leq l \\
      \omega_\delta|_l-\frac{\omega_{pu}}{c\sqrt{2}}\frac{m_ec^2\gamma_{b0}}{E_0e}\left(\frac{1}{\sqrt{\bar{\gamma}_b}}-\frac{1}{\sqrt{\gamma_{bl}}}\right) & l< s\leq l+L \\
      \omega_\delta|_{l+L}+\frac{1}{2}\frac{\omega_{pu}}{c\sqrt{2\gamma_{bL}}}\frac{\gamma_{b0}}{\gamma_{bL}}G_1^* & l+L< s,
   \end{cases}
\]
\end{widetext}
where $G_1$, $G_2$, $D_1$, and $D_2$ are integrals over the plasma density defined in the text. The starred quantities are the integrals evaluated over only the exit ramp:
\begingroup
\allowdisplaybreaks
\begin{align*}
G_1^* =& G_1(s)-G_1(l+L) \\
G_2^* =& G_2(s)-G_2(l+L) \\
D_1^* =& \int_{l+L}^sds'\sqrt{\eta(s')}G_1^*(s') \\
D_2^* =& \int_{l+L}^sds'\sqrt{\eta(s')}G_2^*(s').
\end{align*}
\endgroup
Combining these expressions with Eqs.~(\ref{eqn:CSLinear}), (\ref{eqn:I1}) and (\ref{eqn:I2}) gives $H_1$, $H_2$, $\psi_1$, and $\psi_2$.  These can be inserting into Eq.~(\ref{eq:emGrowth}) to find the emittance growth and Eqs.~(\ref{eqn:moments})-(\ref{eqn:moments_end}) to find the evolution of the beam moments.

\bibliography{bibliography.bib}

\begin{thebibliography}{44}%
\makeatletter
\providecommand \@ifxundefined [1]{%
 \@ifx{#1\undefined}
}%
\providecommand \@ifnum [1]{%
 \ifnum #1\expandafter \@firstoftwo
 \else \expandafter \@secondoftwo
 \fi
}%
\providecommand \@ifx [1]{%
 \ifx #1\expandafter \@firstoftwo
 \else \expandafter \@secondoftwo
 \fi
}%
\providecommand \natexlab [1]{#1}%
\providecommand \enquote  [1]{``#1''}%
\providecommand \bibnamefont  [1]{#1}%
\providecommand \bibfnamefont [1]{#1}%
\providecommand \citenamefont [1]{#1}%
\providecommand \href@noop [0]{\@secondoftwo}%
\providecommand \href [0]{\begingroup \@sanitize@url \@href}%
\providecommand \@href[1]{\@@startlink{#1}\@@href}%
\providecommand \@@href[1]{\endgroup#1\@@endlink}%
\providecommand \@sanitize@url [0]{\catcode `\\12\catcode `\$12\catcode
  `\&12\catcode `\#12\catcode `\^12\catcode `\_12\catcode `\%12\relax}%
\providecommand \@@startlink[1]{}%
\providecommand \@@endlink[0]{}%
\providecommand \url  [0]{\begingroup\@sanitize@url \@url }%
\providecommand \@url [1]{\endgroup\@href {#1}{\urlprefix }}%
\providecommand \urlprefix  [0]{URL }%
\providecommand \Eprint [0]{\href }%
\providecommand \doibase [0]{http://dx.doi.org/}%
\providecommand \selectlanguage [0]{\@gobble}%
\providecommand \bibinfo  [0]{\@secondoftwo}%
\providecommand \bibfield  [0]{\@secondoftwo}%
\providecommand \translation [1]{[#1]}%
\providecommand \BibitemOpen [0]{}%
\providecommand \bibitemStop [0]{}%
\providecommand \bibitemNoStop [0]{.\EOS\space}%
\providecommand \EOS [0]{\spacefactor3000\relax}%
\providecommand \BibitemShut  [1]{\csname bibitem#1\endcsname}%
\let\auto@bib@innerbib\@empty
\bibitem [{\citenamefont {Blumenfeld}\ \emph {et~al.}(2007)\citenamefont
  {Blumenfeld}, \citenamefont {Clayton}, \citenamefont {Decker}, \citenamefont
  {Hogan}, \citenamefont {Huang}, \citenamefont {Ischebeck}, \citenamefont
  {Iverson}, \citenamefont {Joshi}, \citenamefont {Katsouleas}, \citenamefont
  {Kirby}, \citenamefont {Lu}, \citenamefont {Marsh}, \citenamefont {Mori},
  \citenamefont {Muggli}, \citenamefont {Oz}, \citenamefont {Siemann},
  \citenamefont {Walz},\ and\ \citenamefont {Zhou}}]{Blumenfeld2007}%
  \BibitemOpen
  \bibfield  {author} {\bibinfo {author} {\bibfnamefont {I.}~\bibnamefont
  {Blumenfeld}}, \bibinfo {author} {\bibfnamefont {C.~E.}\ \bibnamefont
  {Clayton}}, \bibinfo {author} {\bibfnamefont {F.-J.}\ \bibnamefont {Decker}},
  \bibinfo {author} {\bibfnamefont {M.~J.}\ \bibnamefont {Hogan}}, \bibinfo
  {author} {\bibfnamefont {C.}~\bibnamefont {Huang}}, \bibinfo {author}
  {\bibfnamefont {R.}~\bibnamefont {Ischebeck}}, \bibinfo {author}
  {\bibfnamefont {R.}~\bibnamefont {Iverson}}, \bibinfo {author} {\bibfnamefont
  {C.}~\bibnamefont {Joshi}}, \bibinfo {author} {\bibfnamefont
  {T.}~\bibnamefont {Katsouleas}}, \bibinfo {author} {\bibfnamefont
  {N.}~\bibnamefont {Kirby}}, \bibinfo {author} {\bibfnamefont
  {W.}~\bibnamefont {Lu}}, \bibinfo {author} {\bibfnamefont {K.~A.}\
  \bibnamefont {Marsh}}, \bibinfo {author} {\bibfnamefont {W.~B.}\ \bibnamefont
  {Mori}}, \bibinfo {author} {\bibfnamefont {P.}~\bibnamefont {Muggli}},
  \bibinfo {author} {\bibfnamefont {E.}~\bibnamefont {Oz}}, \bibinfo {author}
  {\bibfnamefont {R.~H.}\ \bibnamefont {Siemann}}, \bibinfo {author}
  {\bibfnamefont {D.}~\bibnamefont {Walz}}, \ and\ \bibinfo {author}
  {\bibfnamefont {M.}~\bibnamefont {Zhou}},\ }\href {\doibase
  10.1038/nature05538} {\bibfield  {journal} {\bibinfo  {journal} {Nature}\
  }\textbf {\bibinfo {volume} {445}},\ \bibinfo {pages} {741} (\bibinfo {year}
  {2007})}\BibitemShut {NoStop}%
\bibitem [{\citenamefont {Litos}\ \emph {et~al.}(2014)\citenamefont {Litos},
  \citenamefont {Adli}, \citenamefont {An}, \citenamefont {Clarke},
  \citenamefont {Clayton}, \citenamefont {Corde}, \citenamefont {Delahaye},
  \citenamefont {England}, \citenamefont {Fisher}, \citenamefont {Frederico},
  \citenamefont {Gessner}, \citenamefont {Green}, \citenamefont {Hogan},
  \citenamefont {Joshi}, \citenamefont {Lu}, \citenamefont {Marsh},
  \citenamefont {Mori}, \citenamefont {Muggli}, \citenamefont
  {Vafaei-Najafabadi}, \citenamefont {Walz}, \citenamefont {White},
  \citenamefont {Wu}, \citenamefont {Yakimenko},\ and\ \citenamefont
  {Yocky}}]{Litos2014}%
  \BibitemOpen
  \bibfield  {author} {\bibinfo {author} {\bibfnamefont {M.}~\bibnamefont
  {Litos}}, \bibinfo {author} {\bibfnamefont {E.}~\bibnamefont {Adli}},
  \bibinfo {author} {\bibfnamefont {W.}~\bibnamefont {An}}, \bibinfo {author}
  {\bibfnamefont {C.~I.}\ \bibnamefont {Clarke}}, \bibinfo {author}
  {\bibfnamefont {C.~E.}\ \bibnamefont {Clayton}}, \bibinfo {author}
  {\bibfnamefont {S.}~\bibnamefont {Corde}}, \bibinfo {author} {\bibfnamefont
  {J.~P.}\ \bibnamefont {Delahaye}}, \bibinfo {author} {\bibfnamefont {R.~J.}\
  \bibnamefont {England}}, \bibinfo {author} {\bibfnamefont {A.~S.}\
  \bibnamefont {Fisher}}, \bibinfo {author} {\bibfnamefont {J.}~\bibnamefont
  {Frederico}}, \bibinfo {author} {\bibfnamefont {S.}~\bibnamefont {Gessner}},
  \bibinfo {author} {\bibfnamefont {S.~Z.}\ \bibnamefont {Green}}, \bibinfo
  {author} {\bibfnamefont {M.~J.}\ \bibnamefont {Hogan}}, \bibinfo {author}
  {\bibfnamefont {C.}~\bibnamefont {Joshi}}, \bibinfo {author} {\bibfnamefont
  {W.}~\bibnamefont {Lu}}, \bibinfo {author} {\bibfnamefont {K.~A.}\
  \bibnamefont {Marsh}}, \bibinfo {author} {\bibfnamefont {W.~B.}\ \bibnamefont
  {Mori}}, \bibinfo {author} {\bibfnamefont {P.}~\bibnamefont {Muggli}},
  \bibinfo {author} {\bibfnamefont {N.}~\bibnamefont {Vafaei-Najafabadi}},
  \bibinfo {author} {\bibfnamefont {D.}~\bibnamefont {Walz}}, \bibinfo {author}
  {\bibfnamefont {G.}~\bibnamefont {White}}, \bibinfo {author} {\bibfnamefont
  {Z.}~\bibnamefont {Wu}}, \bibinfo {author} {\bibfnamefont {V.}~\bibnamefont
  {Yakimenko}}, \ and\ \bibinfo {author} {\bibfnamefont {G.}~\bibnamefont
  {Yocky}},\ }\href {\doibase 10.1038/nature13882} {\bibfield  {journal}
  {\bibinfo  {journal} {Nature}\ }\textbf {\bibinfo {volume} {515}},\ \bibinfo
  {pages} {92} (\bibinfo {year} {2014})}\BibitemShut {NoStop}%
\bibitem [{\citenamefont {Litos}\ \emph {et~al.}(2016)\citenamefont {Litos},
  \citenamefont {Adli}, \citenamefont {Allen}, \citenamefont {An},
  \citenamefont {Clarke}, \citenamefont {Corde}, \citenamefont {Clayton},
  \citenamefont {Frederico}, \citenamefont {Gessner}, \citenamefont {Green},
  \citenamefont {Hogan}, \citenamefont {Joshi}, \citenamefont {Lu},
  \citenamefont {Marsh}, \citenamefont {Mori}, \citenamefont {Schmeltz},
  \citenamefont {Vafaei-Najafabadi},\ and\ \citenamefont
  {Yakimenko}}]{Litos2016}%
  \BibitemOpen
  \bibfield  {author} {\bibinfo {author} {\bibfnamefont {M.}~\bibnamefont
  {Litos}}, \bibinfo {author} {\bibfnamefont {E.}~\bibnamefont {Adli}},
  \bibinfo {author} {\bibfnamefont {J.~M.}\ \bibnamefont {Allen}}, \bibinfo
  {author} {\bibfnamefont {W.}~\bibnamefont {An}}, \bibinfo {author}
  {\bibfnamefont {C.~I.}\ \bibnamefont {Clarke}}, \bibinfo {author}
  {\bibfnamefont {S.}~\bibnamefont {Corde}}, \bibinfo {author} {\bibfnamefont
  {C.~E.}\ \bibnamefont {Clayton}}, \bibinfo {author} {\bibfnamefont
  {J.}~\bibnamefont {Frederico}}, \bibinfo {author} {\bibfnamefont {S.~J.}\
  \bibnamefont {Gessner}}, \bibinfo {author} {\bibfnamefont {S.~Z.}\
  \bibnamefont {Green}}, \bibinfo {author} {\bibfnamefont {M.~J.}\ \bibnamefont
  {Hogan}}, \bibinfo {author} {\bibfnamefont {C.}~\bibnamefont {Joshi}},
  \bibinfo {author} {\bibfnamefont {W.}~\bibnamefont {Lu}}, \bibinfo {author}
  {\bibfnamefont {K.~A.}\ \bibnamefont {Marsh}}, \bibinfo {author}
  {\bibfnamefont {W.~B.}\ \bibnamefont {Mori}}, \bibinfo {author}
  {\bibfnamefont {M.}~\bibnamefont {Schmeltz}}, \bibinfo {author}
  {\bibfnamefont {N.}~\bibnamefont {Vafaei-Najafabadi}}, \ and\ \bibinfo
  {author} {\bibfnamefont {V.}~\bibnamefont {Yakimenko}},\ }\href {\doibase
  10.1088/0741-3335/58/3/034017} {\bibfield  {journal} {\bibinfo  {journal}
  {Plasma Physics and Controlled Fusion}\ }\textbf {\bibinfo {volume} {58}},\
  \bibinfo {pages} {034017} (\bibinfo {year} {2016})},\ \Eprint
  {http://arxiv.org/abs/1511.06743} {arXiv:1511.06743} \BibitemShut {NoStop}%
\bibitem [{\citenamefont {Michel}\ \emph {et~al.}(2006)\citenamefont {Michel},
  \citenamefont {Schroeder}, \citenamefont {Shadwick}, \citenamefont {Esarey},\
  and\ \citenamefont {Leemans}}]{Michel2006}%
  \BibitemOpen
  \bibfield  {author} {\bibinfo {author} {\bibfnamefont {P.}~\bibnamefont
  {Michel}}, \bibinfo {author} {\bibfnamefont {C.~B.}\ \bibnamefont
  {Schroeder}}, \bibinfo {author} {\bibfnamefont {B.~A.}\ \bibnamefont
  {Shadwick}}, \bibinfo {author} {\bibfnamefont {E.}~\bibnamefont {Esarey}}, \
  and\ \bibinfo {author} {\bibfnamefont {W.~P.}\ \bibnamefont {Leemans}},\
  }\href {\doibase 10.1103/PhysRevE.74.026501} {\bibfield  {journal} {\bibinfo
  {journal} {Physical Review E - Statistical, Nonlinear, and Soft Matter
  Physics}\ }\textbf {\bibinfo {volume} {74}},\ \bibinfo {pages} {026501}
  (\bibinfo {year} {2006})}\BibitemShut {NoStop}%
\bibitem [{\citenamefont {Mehrling}\ \emph {et~al.}(2012)\citenamefont
  {Mehrling}, \citenamefont {Grebenyuk}, \citenamefont {Tsung}, \citenamefont
  {Floettmann},\ and\ \citenamefont {Osterhoff}}]{Mehrling2012}%
  \BibitemOpen
  \bibfield  {author} {\bibinfo {author} {\bibfnamefont {T.}~\bibnamefont
  {Mehrling}}, \bibinfo {author} {\bibfnamefont {J.}~\bibnamefont {Grebenyuk}},
  \bibinfo {author} {\bibfnamefont {F.~S.}\ \bibnamefont {Tsung}}, \bibinfo
  {author} {\bibfnamefont {K.}~\bibnamefont {Floettmann}}, \ and\ \bibinfo
  {author} {\bibfnamefont {J.}~\bibnamefont {Osterhoff}},\ }\href {\doibase
  10.1103/PhysRevSTAB.15.111303} {\bibfield  {journal} {\bibinfo  {journal}
  {Physical Review Special Topics - Accelerators and Beams}\ }\textbf {\bibinfo
  {volume} {15}},\ \bibinfo {pages} {111303} (\bibinfo {year}
  {2012})}\BibitemShut {NoStop}%
\bibitem [{\citenamefont {Floettmann}(2014)}]{Floettmann2014}%
  \BibitemOpen
  \bibfield  {author} {\bibinfo {author} {\bibfnamefont {K.}~\bibnamefont
  {Floettmann}},\ }\href {\doibase 10.1103/PhysRevSTAB.17.054402} {\bibfield
  {journal} {\bibinfo  {journal} {Physical Review Special Topics - Accelerators
  and Beams}\ }\textbf {\bibinfo {volume} {17}},\ \bibinfo {pages} {054402}
  (\bibinfo {year} {2014})}\BibitemShut {NoStop}%
\bibitem [{\citenamefont {Dornmair}\ \emph {et~al.}(2015)\citenamefont
  {Dornmair}, \citenamefont {Floettmann},\ and\ \citenamefont
  {Maier}}]{Dornmair2015}%
  \BibitemOpen
  \bibfield  {author} {\bibinfo {author} {\bibfnamefont {I.}~\bibnamefont
  {Dornmair}}, \bibinfo {author} {\bibfnamefont {K.}~\bibnamefont
  {Floettmann}}, \ and\ \bibinfo {author} {\bibfnamefont {A.~R.}\ \bibnamefont
  {Maier}},\ }\href {\doibase 10.1103/PhysRevSTAB.18.041302} {\bibfield
  {journal} {\bibinfo  {journal} {Physical Review Special Topics - Accelerators
  and Beams}\ }\textbf {\bibinfo {volume} {18}},\ \bibinfo {pages} {041302}
  (\bibinfo {year} {2015})}\BibitemShut {NoStop}%
\bibitem [{\citenamefont {Xu}\ \emph {et~al.}(2016)\citenamefont {Xu},
  \citenamefont {Hua}, \citenamefont {Wu}, \citenamefont {Zhang}, \citenamefont
  {Li}, \citenamefont {Wan}, \citenamefont {Pai}, \citenamefont {Lu},
  \citenamefont {An}, \citenamefont {Yu}, \citenamefont {Hogan}, \citenamefont
  {Joshi},\ and\ \citenamefont {Mori}}]{Xu2016}%
  \BibitemOpen
  \bibfield  {author} {\bibinfo {author} {\bibfnamefont {X.~L.}\ \bibnamefont
  {Xu}}, \bibinfo {author} {\bibfnamefont {J.~F.}\ \bibnamefont {Hua}},
  \bibinfo {author} {\bibfnamefont {Y.~P.}\ \bibnamefont {Wu}}, \bibinfo
  {author} {\bibfnamefont {C.~J.}\ \bibnamefont {Zhang}}, \bibinfo {author}
  {\bibfnamefont {F.}~\bibnamefont {Li}}, \bibinfo {author} {\bibfnamefont
  {Y.}~\bibnamefont {Wan}}, \bibinfo {author} {\bibfnamefont {C.-H.}\
  \bibnamefont {Pai}}, \bibinfo {author} {\bibfnamefont {W.}~\bibnamefont
  {Lu}}, \bibinfo {author} {\bibfnamefont {W.}~\bibnamefont {An}}, \bibinfo
  {author} {\bibfnamefont {P.}~\bibnamefont {Yu}}, \bibinfo {author}
  {\bibfnamefont {M.~J.}\ \bibnamefont {Hogan}}, \bibinfo {author}
  {\bibfnamefont {C.}~\bibnamefont {Joshi}}, \ and\ \bibinfo {author}
  {\bibfnamefont {W.~B.}\ \bibnamefont {Mori}},\ }\href {\doibase
  10.1103/PhysRevLett.116.124801} {\bibfield  {journal} {\bibinfo  {journal}
  {Physical Review Letters}\ }\textbf {\bibinfo {volume} {116}},\ \bibinfo
  {pages} {124801} (\bibinfo {year} {2016})},\ \Eprint
  {http://arxiv.org/abs/1411.4386} {arXiv:1411.4386} \BibitemShut {NoStop}%
\bibitem [{\citenamefont {Aschikhin}\ \emph {et~al.}(2018)\citenamefont
  {Aschikhin}, \citenamefont {Mehrling}, \citenamefont {{Martinez de la
  Ossa}},\ and\ \citenamefont {Osterhoff}}]{Aschikhin2018}%
  \BibitemOpen
  \bibfield  {author} {\bibinfo {author} {\bibfnamefont {A.}~\bibnamefont
  {Aschikhin}}, \bibinfo {author} {\bibfnamefont {T.~J.}\ \bibnamefont
  {Mehrling}}, \bibinfo {author} {\bibfnamefont {A.}~\bibnamefont {{Martinez de
  la Ossa}}}, \ and\ \bibinfo {author} {\bibfnamefont {J.}~\bibnamefont
  {Osterhoff}},\ }\href {\doibase 10.1016/j.nima.2018.02.065} {\bibfield
  {journal} {\bibinfo  {journal} {Nuclear Instruments and Methods in Physics
  Research, Section A: Accelerators, Spectrometers, Detectors and Associated
  Equipment}\ }\textbf {\bibinfo {volume} {909}},\ \bibinfo {pages} {414}
  (\bibinfo {year} {2018})},\ \Eprint {http://arxiv.org/abs/1802.03968}
  {arXiv:1802.03968} \BibitemShut {NoStop}%
\bibitem [{\citenamefont {Ariniello}\ \emph {et~al.}(2019)\citenamefont
  {Ariniello}, \citenamefont {Doss}, \citenamefont {Hunt-Stone}, \citenamefont
  {Cary},\ and\ \citenamefont {Litos}}]{ariniello:2019prab}%
  \BibitemOpen
  \bibfield  {author} {\bibinfo {author} {\bibfnamefont {R.}~\bibnamefont
  {Ariniello}}, \bibinfo {author} {\bibfnamefont {C.~E.}\ \bibnamefont {Doss}},
  \bibinfo {author} {\bibfnamefont {K.}~\bibnamefont {Hunt-Stone}}, \bibinfo
  {author} {\bibfnamefont {J.~R.}\ \bibnamefont {Cary}}, \ and\ \bibinfo
  {author} {\bibfnamefont {M.~D.}\ \bibnamefont {Litos}},\ }\href {\doibase
  10.1103/PhysRevAccelBeams.22.041304} {\bibfield  {journal} {\bibinfo
  {journal} {Physical Review Accelerators and Beams}\ }\textbf {\bibinfo
  {volume} {22}},\ \bibinfo {pages} {041304} (\bibinfo {year}
  {2019})}\BibitemShut {NoStop}%
\bibitem [{\citenamefont {Zhao}\ \emph {et~al.}(2020)\citenamefont {Zhao},
  \citenamefont {An}, \citenamefont {Xu}, \citenamefont {Li}, \citenamefont
  {Hildebrand}, \citenamefont {Hogan}, \citenamefont {Yakimenko}, \citenamefont
  {Joshi},\ and\ \citenamefont {Mori}}]{Zhao2020}%
  \BibitemOpen
  \bibfield  {author} {\bibinfo {author} {\bibfnamefont {Y.}~\bibnamefont
  {Zhao}}, \bibinfo {author} {\bibfnamefont {W.}~\bibnamefont {An}}, \bibinfo
  {author} {\bibfnamefont {X.}~\bibnamefont {Xu}}, \bibinfo {author}
  {\bibfnamefont {F.}~\bibnamefont {Li}}, \bibinfo {author} {\bibfnamefont
  {L.}~\bibnamefont {Hildebrand}}, \bibinfo {author} {\bibfnamefont {M.~J.}\
  \bibnamefont {Hogan}}, \bibinfo {author} {\bibfnamefont {V.}~\bibnamefont
  {Yakimenko}}, \bibinfo {author} {\bibfnamefont {C.}~\bibnamefont {Joshi}}, \
  and\ \bibinfo {author} {\bibfnamefont {W.~B.}\ \bibnamefont {Mori}},\ }\href
  {\doibase 10.1103/PhysRevAccelBeams.23.011302} {\bibfield  {journal}
  {\bibinfo  {journal} {Physical Review Accelerators and Beams}\ }\textbf
  {\bibinfo {volume} {23}},\ \bibinfo {pages} {011302} (\bibinfo {year}
  {2020})}\BibitemShut {NoStop}%
\bibitem [{\citenamefont {Whittum}\ \emph {et~al.}(1991)\citenamefont
  {Whittum}, \citenamefont {Sharp}, \citenamefont {Yu}, \citenamefont {Lampe},\
  and\ \citenamefont {Joyce}}]{Whittum1991}%
  \BibitemOpen
  \bibfield  {author} {\bibinfo {author} {\bibfnamefont {D.~H.}\ \bibnamefont
  {Whittum}}, \bibinfo {author} {\bibfnamefont {W.~M.}\ \bibnamefont {Sharp}},
  \bibinfo {author} {\bibfnamefont {S.~S.}\ \bibnamefont {Yu}}, \bibinfo
  {author} {\bibfnamefont {M.}~\bibnamefont {Lampe}}, \ and\ \bibinfo {author}
  {\bibfnamefont {G.}~\bibnamefont {Joyce}},\ }\href {\doibase
  10.1103/PhysRevLett.67.991} {\bibfield  {journal} {\bibinfo  {journal}
  {Physical Review Letters}\ }\textbf {\bibinfo {volume} {67}},\ \bibinfo
  {pages} {991} (\bibinfo {year} {1991})}\BibitemShut {NoStop}%
\bibitem [{\citenamefont {Lampe}\ \emph {et~al.}(1993)\citenamefont {Lampe},
  \citenamefont {Joyce}, \citenamefont {Slinker},\ and\ \citenamefont
  {Whittum}}]{Lampe1993}%
  \BibitemOpen
  \bibfield  {author} {\bibinfo {author} {\bibfnamefont {M.}~\bibnamefont
  {Lampe}}, \bibinfo {author} {\bibfnamefont {G.}~\bibnamefont {Joyce}},
  \bibinfo {author} {\bibfnamefont {S.~P.}\ \bibnamefont {Slinker}}, \ and\
  \bibinfo {author} {\bibfnamefont {D.~H.}\ \bibnamefont {Whittum}},\ }\href
  {\doibase 10.1063/1.860772} {\bibfield  {journal} {\bibinfo  {journal}
  {Physics of Fluids B}\ }\textbf {\bibinfo {volume} {5}},\ \bibinfo {pages}
  {1888} (\bibinfo {year} {1993})}\BibitemShut {NoStop}%
\bibitem [{\citenamefont {Geraci}\ and\ \citenamefont
  {Whittum}(2000)}]{Geraci2000b}%
  \BibitemOpen
  \bibfield  {author} {\bibinfo {author} {\bibfnamefont {A.~A.}\ \bibnamefont
  {Geraci}}\ and\ \bibinfo {author} {\bibfnamefont {D.~H.}\ \bibnamefont
  {Whittum}},\ }\href {\doibase 10.1063/1.874207} {\bibfield  {journal}
  {\bibinfo  {journal} {Physics of Plasmas}\ }\textbf {\bibinfo {volume} {7}},\
  \bibinfo {pages} {3431} (\bibinfo {year} {2000})}\BibitemShut {NoStop}%
\bibitem [{\citenamefont {Deng}\ \emph {et~al.}(2006)\citenamefont {Deng},
  \citenamefont {Barnes}, \citenamefont {Clayton}, \citenamefont {O'Connell},
  \citenamefont {Decker}, \citenamefont {Fonseca}, \citenamefont {Huang},
  \citenamefont {Hogan}, \citenamefont {Iverson}, \citenamefont {Johnson},
  \citenamefont {Joshi}, \citenamefont {Katsouleas}, \citenamefont {Krejcik},
  \citenamefont {Lu}, \citenamefont {Mori}, \citenamefont {Muggli},
  \citenamefont {Oz}, \citenamefont {Tsung}, \citenamefont {Walz},\ and\
  \citenamefont {Zhou}}]{Deng2006b}%
  \BibitemOpen
  \bibfield  {author} {\bibinfo {author} {\bibfnamefont {S.}~\bibnamefont
  {Deng}}, \bibinfo {author} {\bibfnamefont {C.~D.}\ \bibnamefont {Barnes}},
  \bibinfo {author} {\bibfnamefont {C.~E.}\ \bibnamefont {Clayton}}, \bibinfo
  {author} {\bibfnamefont {C.}~\bibnamefont {O'Connell}}, \bibinfo {author}
  {\bibfnamefont {F.~J.}\ \bibnamefont {Decker}}, \bibinfo {author}
  {\bibfnamefont {R.~A.}\ \bibnamefont {Fonseca}}, \bibinfo {author}
  {\bibfnamefont {C.}~\bibnamefont {Huang}}, \bibinfo {author} {\bibfnamefont
  {M.~J.}\ \bibnamefont {Hogan}}, \bibinfo {author} {\bibfnamefont
  {R.}~\bibnamefont {Iverson}}, \bibinfo {author} {\bibfnamefont {D.~K.}\
  \bibnamefont {Johnson}}, \bibinfo {author} {\bibfnamefont {C.}~\bibnamefont
  {Joshi}}, \bibinfo {author} {\bibfnamefont {T.}~\bibnamefont {Katsouleas}},
  \bibinfo {author} {\bibfnamefont {P.}~\bibnamefont {Krejcik}}, \bibinfo
  {author} {\bibfnamefont {W.}~\bibnamefont {Lu}}, \bibinfo {author}
  {\bibfnamefont {W.~B.}\ \bibnamefont {Mori}}, \bibinfo {author}
  {\bibfnamefont {P.}~\bibnamefont {Muggli}}, \bibinfo {author} {\bibfnamefont
  {E.}~\bibnamefont {Oz}}, \bibinfo {author} {\bibfnamefont {F.}~\bibnamefont
  {Tsung}}, \bibinfo {author} {\bibfnamefont {D.}~\bibnamefont {Walz}}, \ and\
  \bibinfo {author} {\bibfnamefont {M.}~\bibnamefont {Zhou}},\ }\href {\doibase
  10.1103/PhysRevLett.96.045001} {\bibfield  {journal} {\bibinfo  {journal}
  {Physical Review Letters}\ }\textbf {\bibinfo {volume} {96}},\ \bibinfo
  {pages} {045001} (\bibinfo {year} {2006})}\BibitemShut {NoStop}%
\bibitem [{\citenamefont {Huang}\ \emph {et~al.}(2007)\citenamefont {Huang},
  \citenamefont {Lu}, \citenamefont {Zhou}, \citenamefont {Clayton},
  \citenamefont {Joshi}, \citenamefont {Mori}, \citenamefont {Muggli},
  \citenamefont {Deng}, \citenamefont {Oz}, \citenamefont {Katsouleas},
  \citenamefont {Hogan}, \citenamefont {Blumenfeld}, \citenamefont {Decker},
  \citenamefont {Ischebeck}, \citenamefont {Iverson}, \citenamefont {Kirby},\
  and\ \citenamefont {Walz}}]{Huang2007}%
  \BibitemOpen
  \bibfield  {author} {\bibinfo {author} {\bibfnamefont {C.}~\bibnamefont
  {Huang}}, \bibinfo {author} {\bibfnamefont {W.}~\bibnamefont {Lu}}, \bibinfo
  {author} {\bibfnamefont {M.}~\bibnamefont {Zhou}}, \bibinfo {author}
  {\bibfnamefont {C.~E.}\ \bibnamefont {Clayton}}, \bibinfo {author}
  {\bibfnamefont {C.}~\bibnamefont {Joshi}}, \bibinfo {author} {\bibfnamefont
  {W.~B.}\ \bibnamefont {Mori}}, \bibinfo {author} {\bibfnamefont
  {P.}~\bibnamefont {Muggli}}, \bibinfo {author} {\bibfnamefont
  {S.}~\bibnamefont {Deng}}, \bibinfo {author} {\bibfnamefont {E.}~\bibnamefont
  {Oz}}, \bibinfo {author} {\bibfnamefont {T.}~\bibnamefont {Katsouleas}},
  \bibinfo {author} {\bibfnamefont {M.~J.}\ \bibnamefont {Hogan}}, \bibinfo
  {author} {\bibfnamefont {I.}~\bibnamefont {Blumenfeld}}, \bibinfo {author}
  {\bibfnamefont {F.~J.}\ \bibnamefont {Decker}}, \bibinfo {author}
  {\bibfnamefont {R.}~\bibnamefont {Ischebeck}}, \bibinfo {author}
  {\bibfnamefont {R.~H.}\ \bibnamefont {Iverson}}, \bibinfo {author}
  {\bibfnamefont {N.~A.}\ \bibnamefont {Kirby}}, \ and\ \bibinfo {author}
  {\bibfnamefont {D.}~\bibnamefont {Walz}},\ }\href {\doibase
  10.1103/PhysRevLett.99.255001} {\bibfield  {journal} {\bibinfo  {journal}
  {Physical Review Letters}\ }\textbf {\bibinfo {volume} {99}},\ \bibinfo
  {pages} {255001} (\bibinfo {year} {2007})}\BibitemShut {NoStop}%
\bibitem [{\citenamefont {Mehrling}\ \emph {et~al.}(2017)\citenamefont
  {Mehrling}, \citenamefont {Fonseca}, \citenamefont {{Martinez de la Ossa}},\
  and\ \citenamefont {Vieira}}]{Mehrling2017}%
  \BibitemOpen
  \bibfield  {author} {\bibinfo {author} {\bibfnamefont {T.~J.}\ \bibnamefont
  {Mehrling}}, \bibinfo {author} {\bibfnamefont {R.~A.}\ \bibnamefont
  {Fonseca}}, \bibinfo {author} {\bibfnamefont {A.}~\bibnamefont {{Martinez de
  la Ossa}}}, \ and\ \bibinfo {author} {\bibfnamefont {J.}~\bibnamefont
  {Vieira}},\ }\href {\doibase 10.1103/PhysRevLett.118.174801} {\bibfield
  {journal} {\bibinfo  {journal} {Physical Review Letters}\ }\textbf {\bibinfo
  {volume} {118}},\ \bibinfo {pages} {174801} (\bibinfo {year}
  {2017})}\BibitemShut {NoStop}%
\bibitem [{\citenamefont {Lebedev}\ \emph {et~al.}(2017)\citenamefont
  {Lebedev}, \citenamefont {Burov},\ and\ \citenamefont
  {Nagaitsev}}]{Lebedev2017}%
  \BibitemOpen
  \bibfield  {author} {\bibinfo {author} {\bibfnamefont {V.}~\bibnamefont
  {Lebedev}}, \bibinfo {author} {\bibfnamefont {A.}~\bibnamefont {Burov}}, \
  and\ \bibinfo {author} {\bibfnamefont {S.}~\bibnamefont {Nagaitsev}},\ }\href
  {\doibase 10.1103/PhysRevAccelBeams.20.121301} {\bibfield  {journal}
  {\bibinfo  {journal} {Physical Review Accelerators and Beams}\ }\textbf
  {\bibinfo {volume} {20}},\ \bibinfo {pages} {121301} (\bibinfo {year}
  {2017})}\BibitemShut {NoStop}%
\bibitem [{\citenamefont {Mehrling}\ \emph
  {et~al.}(2018{\natexlab{a}})\citenamefont {Mehrling}, \citenamefont
  {Benedetti}, \citenamefont {Schroeder}, \citenamefont {Esarey},\ and\
  \citenamefont {Leemans}}]{Mehrling2018}%
  \BibitemOpen
  \bibfield  {author} {\bibinfo {author} {\bibfnamefont {T.~J.}\ \bibnamefont
  {Mehrling}}, \bibinfo {author} {\bibfnamefont {C.}~\bibnamefont {Benedetti}},
  \bibinfo {author} {\bibfnamefont {C.~B.}\ \bibnamefont {Schroeder}}, \bibinfo
  {author} {\bibfnamefont {E.}~\bibnamefont {Esarey}}, \ and\ \bibinfo {author}
  {\bibfnamefont {W.~P.}\ \bibnamefont {Leemans}},\ }\href {\doibase
  10.1103/PhysRevLett.121.264802} {\bibfield  {journal} {\bibinfo  {journal}
  {Physical Review Letters}\ }\textbf {\bibinfo {volume} {121}},\ \bibinfo
  {pages} {264802} (\bibinfo {year} {2018}{\natexlab{a}})}\BibitemShut
  {NoStop}%
\bibitem [{\citenamefont {Mehrling}\ \emph
  {et~al.}(2018{\natexlab{b}})\citenamefont {Mehrling}, \citenamefont
  {Benedetti}, \citenamefont {Schroeder}, \citenamefont {{Martinez de la
  Ossa}}, \citenamefont {Osterhoff}, \citenamefont {Esarey},\ and\
  \citenamefont {Leemans}}]{Mehrling2018a}%
  \BibitemOpen
  \bibfield  {author} {\bibinfo {author} {\bibfnamefont {T.~J.}\ \bibnamefont
  {Mehrling}}, \bibinfo {author} {\bibfnamefont {C.}~\bibnamefont {Benedetti}},
  \bibinfo {author} {\bibfnamefont {C.~B.}\ \bibnamefont {Schroeder}}, \bibinfo
  {author} {\bibfnamefont {A.}~\bibnamefont {{Martinez de la Ossa}}}, \bibinfo
  {author} {\bibfnamefont {J.}~\bibnamefont {Osterhoff}}, \bibinfo {author}
  {\bibfnamefont {E.}~\bibnamefont {Esarey}}, \ and\ \bibinfo {author}
  {\bibfnamefont {W.~P.}\ \bibnamefont {Leemans}},\ }\href {\doibase
  10.1063/1.5017960} {\bibfield  {journal} {\bibinfo  {journal} {Physics of
  Plasmas}\ }\textbf {\bibinfo {volume} {25}},\ \bibinfo {pages} {056703}
  (\bibinfo {year} {2018}{\natexlab{b}})}\BibitemShut {NoStop}%
\bibitem [{\citenamefont {Mehrling}\ \emph {et~al.}(2019)\citenamefont
  {Mehrling}, \citenamefont {Fonseca}, \citenamefont {{Martinez de la Ossa}},\
  and\ \citenamefont {Vieira}}]{Mehrling2019}%
  \BibitemOpen
  \bibfield  {author} {\bibinfo {author} {\bibfnamefont {T.~J.}\ \bibnamefont
  {Mehrling}}, \bibinfo {author} {\bibfnamefont {R.~A.}\ \bibnamefont
  {Fonseca}}, \bibinfo {author} {\bibfnamefont {A.}~\bibnamefont {{Martinez de
  la Ossa}}}, \ and\ \bibinfo {author} {\bibfnamefont {J.}~\bibnamefont
  {Vieira}},\ }\href {\doibase 10.1103/PhysRevAccelBeams.22.031302} {\bibfield
  {journal} {\bibinfo  {journal} {Physical Review Accelerators and Beams}\
  }\textbf {\bibinfo {volume} {22}},\ \bibinfo {pages} {031302} (\bibinfo
  {year} {2019})}\BibitemShut {NoStop}%
\bibitem [{\citenamefont {Assmann}\ and\ \citenamefont
  {Yokoya}(1998)}]{Assmann1998}%
  \BibitemOpen
  \bibfield  {author} {\bibinfo {author} {\bibfnamefont {R.}~\bibnamefont
  {Assmann}}\ and\ \bibinfo {author} {\bibfnamefont {K.}~\bibnamefont
  {Yokoya}},\ }\href {\doibase 10.1016/S0168-9002(98)00187-9} {\bibfield
  {journal} {\bibinfo  {journal} {Nuclear Instruments and Methods in Physics
  Research, Section A: Accelerators, Spectrometers, Detectors and Associated
  Equipment}\ }\textbf {\bibinfo {volume} {410}},\ \bibinfo {pages} {544}
  (\bibinfo {year} {1998})}\BibitemShut {NoStop}%
\bibitem [{\citenamefont {Lindstr{\o}m}\ \emph {et~al.}(2016)\citenamefont
  {Lindstr{\o}m}, \citenamefont {Adli}, \citenamefont {Pfingstner},
  \citenamefont {Marin},\ and\ \citenamefont {Schulte}}]{Lindstrom2016b}%
  \BibitemOpen
  \bibfield  {author} {\bibinfo {author} {\bibfnamefont {C.}~\bibnamefont
  {Lindstr{\o}m}}, \bibinfo {author} {\bibfnamefont {E.}~\bibnamefont {Adli}},
  \bibinfo {author} {\bibfnamefont {J.}~\bibnamefont {Pfingstner}}, \bibinfo
  {author} {\bibfnamefont {E.}~\bibnamefont {Marin}}, \ and\ \bibinfo {author}
  {\bibfnamefont {D.}~\bibnamefont {Schulte}},\ }\href@noop {} {\bibfield
  {journal} {\bibinfo  {journal} {IPAC 2016 - Proceedings of the 7th
  International Particle Accelerator Conference}\ } (\bibinfo {year}
  {2016})}\BibitemShut {NoStop}%
\bibitem [{\citenamefont {Thevenet}\ \emph {et~al.}(2019)\citenamefont
  {Thevenet}, \citenamefont {Lehe}, \citenamefont {Schroeder}, \citenamefont
  {Benedetti}, \citenamefont {Vay}, \citenamefont {Esarey},\ and\ \citenamefont
  {Leemans}}]{Thevenet2019}%
  \BibitemOpen
  \bibfield  {author} {\bibinfo {author} {\bibfnamefont {M.}~\bibnamefont
  {Thevenet}}, \bibinfo {author} {\bibfnamefont {R.}~\bibnamefont {Lehe}},
  \bibinfo {author} {\bibfnamefont {C.~B.}\ \bibnamefont {Schroeder}}, \bibinfo
  {author} {\bibfnamefont {C.}~\bibnamefont {Benedetti}}, \bibinfo {author}
  {\bibfnamefont {J.-L.}\ \bibnamefont {Vay}}, \bibinfo {author} {\bibfnamefont
  {E.}~\bibnamefont {Esarey}}, \ and\ \bibinfo {author} {\bibfnamefont {W.~P.}\
  \bibnamefont {Leemans}},\ }\href {\doibase
  10.1103/PhysRevAccelBeams.22.051302} {\bibfield  {journal} {\bibinfo
  {journal} {Physical Review Accelerators and Beams}\ }\textbf {\bibinfo
  {volume} {22}},\ \bibinfo {pages} {051302} (\bibinfo {year}
  {2019})}\BibitemShut {NoStop}%
\bibitem [{\citenamefont {Raubenheimer}(2000)}]{Raubenheimer2000}%
  \BibitemOpen
  \bibfield  {author} {\bibinfo {author} {\bibfnamefont {T.~O.}\ \bibnamefont
  {Raubenheimer}},\ }\href {\doibase 10.1103/physrevstab.3.121002} {\bibfield
  {journal} {\bibinfo  {journal} {Physical Review Special Topics - Accelerators
  and Beams}\ }\textbf {\bibinfo {volume} {3}},\ \bibinfo {pages} {121002}
  (\bibinfo {year} {2000})}\BibitemShut {NoStop}%
\bibitem [{\citenamefont {Xu}\ \emph {et~al.}(2017)\citenamefont {Xu},
  \citenamefont {Li}, \citenamefont {An}, \citenamefont {Dalichaouch},
  \citenamefont {Yu}, \citenamefont {Lu}, \citenamefont {Joshi},\ and\
  \citenamefont {Mori}}]{Xu2017}%
  \BibitemOpen
  \bibfield  {author} {\bibinfo {author} {\bibfnamefont {X.~L.}\ \bibnamefont
  {Xu}}, \bibinfo {author} {\bibfnamefont {F.}~\bibnamefont {Li}}, \bibinfo
  {author} {\bibfnamefont {W.}~\bibnamefont {An}}, \bibinfo {author}
  {\bibfnamefont {T.~N.}\ \bibnamefont {Dalichaouch}}, \bibinfo {author}
  {\bibfnamefont {P.}~\bibnamefont {Yu}}, \bibinfo {author} {\bibfnamefont
  {W.}~\bibnamefont {Lu}}, \bibinfo {author} {\bibfnamefont {C.}~\bibnamefont
  {Joshi}}, \ and\ \bibinfo {author} {\bibfnamefont {W.~B.}\ \bibnamefont
  {Mori}},\ }\href {\doibase 10.1103/PhysRevAccelBeams.20.111303} {\bibfield
  {journal} {\bibinfo  {journal} {Physical Review Accelerators and Beams}\
  }\textbf {\bibinfo {volume} {20}},\ \bibinfo {pages} {111303} (\bibinfo
  {year} {2017})},\ \Eprint {http://arxiv.org/abs/1610.00788}
  {arXiv:1610.00788} \BibitemShut {NoStop}%
\bibitem [{\citenamefont {Dalichaouch}\ \emph {et~al.}(2020)\citenamefont
  {Dalichaouch}, \citenamefont {Xu}, \citenamefont {Li}, \citenamefont
  {Tableman}, \citenamefont {Tsung}, \citenamefont {An},\ and\ \citenamefont
  {Mori}}]{Dalichaouch2020}%
  \BibitemOpen
  \bibfield  {author} {\bibinfo {author} {\bibfnamefont {T.~N.}\ \bibnamefont
  {Dalichaouch}}, \bibinfo {author} {\bibfnamefont {X.~L.}\ \bibnamefont {Xu}},
  \bibinfo {author} {\bibfnamefont {F.}~\bibnamefont {Li}}, \bibinfo {author}
  {\bibfnamefont {A.}~\bibnamefont {Tableman}}, \bibinfo {author}
  {\bibfnamefont {F.~S.}\ \bibnamefont {Tsung}}, \bibinfo {author}
  {\bibfnamefont {W.}~\bibnamefont {An}}, \ and\ \bibinfo {author}
  {\bibfnamefont {W.~B.}\ \bibnamefont {Mori}},\ }\href {\doibase
  10.1103/physrevaccelbeams.23.021304} {\bibfield  {journal} {\bibinfo
  {journal} {Physical Review Accelerators and Beams}\ }\textbf {\bibinfo
  {volume} {23}},\ \bibinfo {pages} {021304} (\bibinfo {year} {2020})},\
  \Eprint {http://arxiv.org/abs/1909.02689} {arXiv:1909.02689} \BibitemShut
  {NoStop}%
\bibitem [{\citenamefont {Huang}\ \emph {et~al.}(2012)\citenamefont {Huang},
  \citenamefont {Ding},\ and\ \citenamefont {Schroeder}}]{Huang2012}%
  \BibitemOpen
  \bibfield  {author} {\bibinfo {author} {\bibfnamefont {Z.}~\bibnamefont
  {Huang}}, \bibinfo {author} {\bibfnamefont {Y.}~\bibnamefont {Ding}}, \ and\
  \bibinfo {author} {\bibfnamefont {C.~B.}\ \bibnamefont {Schroeder}},\ }\href
  {\doibase 10.1103/PhysRevLett.109.204801} {\bibfield  {journal} {\bibinfo
  {journal} {Physical Review Letters}\ }\textbf {\bibinfo {volume} {109}},\
  \bibinfo {pages} {204801} (\bibinfo {year} {2012})}\BibitemShut {NoStop}%
\bibitem [{\citenamefont {Smith}\ \emph {et~al.}(1979)\citenamefont {Smith},
  \citenamefont {Madey}, \citenamefont {Elias},\ and\ \citenamefont
  {Deacon}}]{Smith1979}%
  \BibitemOpen
  \bibfield  {author} {\bibinfo {author} {\bibfnamefont {T.~I.}\ \bibnamefont
  {Smith}}, \bibinfo {author} {\bibfnamefont {J.~M.~J.}\ \bibnamefont {Madey}},
  \bibinfo {author} {\bibfnamefont {L.~R.}\ \bibnamefont {Elias}}, \ and\
  \bibinfo {author} {\bibfnamefont {D.~A.~G.}\ \bibnamefont {Deacon}},\ }\href
  {\doibase 10.1063/1.326564} {\bibfield  {journal} {\bibinfo  {journal}
  {Journal of Applied Physics}\ }\textbf {\bibinfo {volume} {50}},\ \bibinfo
  {pages} {4580} (\bibinfo {year} {1979})}\BibitemShut {NoStop}%
\bibitem [{\citenamefont {Baxevanis}\ \emph {et~al.}(2014)\citenamefont
  {Baxevanis}, \citenamefont {Ding}, \citenamefont {Huang},\ and\ \citenamefont
  {Ruth}}]{Baxevanis2014}%
  \BibitemOpen
  \bibfield  {author} {\bibinfo {author} {\bibfnamefont {P.}~\bibnamefont
  {Baxevanis}}, \bibinfo {author} {\bibfnamefont {Y.}~\bibnamefont {Ding}},
  \bibinfo {author} {\bibfnamefont {Z.}~\bibnamefont {Huang}}, \ and\ \bibinfo
  {author} {\bibfnamefont {R.}~\bibnamefont {Ruth}},\ }\href {\doibase
  10.1103/PhysRevSTAB.17.020701} {\bibfield  {journal} {\bibinfo  {journal}
  {Physical Review Special Topics - Accelerators and Beams}\ }\textbf {\bibinfo
  {volume} {17}},\ \bibinfo {pages} {020701} (\bibinfo {year}
  {2014})}\BibitemShut {NoStop}%
\bibitem [{\citenamefont {Lu}\ \emph {et~al.}(2006{\natexlab{a}})\citenamefont
  {Lu}, \citenamefont {Huang}, \citenamefont {Zhou}, \citenamefont {Mori},\
  and\ \citenamefont {Katsouleas}}]{Lu2006}%
  \BibitemOpen
  \bibfield  {author} {\bibinfo {author} {\bibfnamefont {W.}~\bibnamefont
  {Lu}}, \bibinfo {author} {\bibfnamefont {C.}~\bibnamefont {Huang}}, \bibinfo
  {author} {\bibfnamefont {M.}~\bibnamefont {Zhou}}, \bibinfo {author}
  {\bibfnamefont {W.~B.}\ \bibnamefont {Mori}}, \ and\ \bibinfo {author}
  {\bibfnamefont {T.}~\bibnamefont {Katsouleas}},\ }\href {\doibase
  10.1103/PhysRevLett.96.165002} {\bibfield  {journal} {\bibinfo  {journal}
  {Physical Review Letters}\ }\textbf {\bibinfo {volume} {96}},\ \bibinfo
  {pages} {1} (\bibinfo {year} {2006}{\natexlab{a}})}\BibitemShut {NoStop}%
\bibitem [{\citenamefont {Lu}\ \emph {et~al.}(2006{\natexlab{b}})\citenamefont
  {Lu}, \citenamefont {Huang}, \citenamefont {Zhou}, \citenamefont {Tzoufras},
  \citenamefont {Tsung}, \citenamefont {Mori},\ and\ \citenamefont
  {Katsouleas}}]{Lu2006a}%
  \BibitemOpen
  \bibfield  {author} {\bibinfo {author} {\bibfnamefont {W.}~\bibnamefont
  {Lu}}, \bibinfo {author} {\bibfnamefont {C.}~\bibnamefont {Huang}}, \bibinfo
  {author} {\bibfnamefont {M.}~\bibnamefont {Zhou}}, \bibinfo {author}
  {\bibfnamefont {M.}~\bibnamefont {Tzoufras}}, \bibinfo {author}
  {\bibfnamefont {F.~S.}\ \bibnamefont {Tsung}}, \bibinfo {author}
  {\bibfnamefont {W.~B.}\ \bibnamefont {Mori}}, \ and\ \bibinfo {author}
  {\bibfnamefont {T.}~\bibnamefont {Katsouleas}},\ }\href {\doibase
  10.1063/1.2203364} {\bibfield  {journal} {\bibinfo  {journal} {Physics of
  Plasmas}\ }\textbf {\bibinfo {volume} {13}},\ \bibinfo {pages} {056709}
  (\bibinfo {year} {2006}{\natexlab{b}})}\BibitemShut {NoStop}%
\bibitem [{\citenamefont {Xu}\ \emph {et~al.}(2014)\citenamefont {Xu},
  \citenamefont {Hua}, \citenamefont {Li}, \citenamefont {Zhang}, \citenamefont
  {Yan}, \citenamefont {Du}, \citenamefont {Huang}, \citenamefont {Chen},
  \citenamefont {Tang}, \citenamefont {Lu}, \citenamefont {Yu}, \citenamefont
  {An}, \citenamefont {Joshi},\ and\ \citenamefont {Mori}}]{Xu2014a}%
  \BibitemOpen
  \bibfield  {author} {\bibinfo {author} {\bibfnamefont {X.~L.}\ \bibnamefont
  {Xu}}, \bibinfo {author} {\bibfnamefont {J.~F.}\ \bibnamefont {Hua}},
  \bibinfo {author} {\bibfnamefont {F.}~\bibnamefont {Li}}, \bibinfo {author}
  {\bibfnamefont {C.~J.}\ \bibnamefont {Zhang}}, \bibinfo {author}
  {\bibfnamefont {L.~X.}\ \bibnamefont {Yan}}, \bibinfo {author} {\bibfnamefont
  {Y.~C.}\ \bibnamefont {Du}}, \bibinfo {author} {\bibfnamefont {W.~H.}\
  \bibnamefont {Huang}}, \bibinfo {author} {\bibfnamefont {H.~B.}\ \bibnamefont
  {Chen}}, \bibinfo {author} {\bibfnamefont {C.~X.}\ \bibnamefont {Tang}},
  \bibinfo {author} {\bibfnamefont {W.}~\bibnamefont {Lu}}, \bibinfo {author}
  {\bibfnamefont {P.}~\bibnamefont {Yu}}, \bibinfo {author} {\bibfnamefont
  {W.}~\bibnamefont {An}}, \bibinfo {author} {\bibfnamefont {C.}~\bibnamefont
  {Joshi}}, \ and\ \bibinfo {author} {\bibfnamefont {W.~B.}\ \bibnamefont
  {Mori}},\ }\href {\doibase 10.1103/PhysRevLett.112.035003} {\bibfield
  {journal} {\bibinfo  {journal} {Physical Review Letters}\ }\textbf {\bibinfo
  {volume} {112}},\ \bibinfo {pages} {035003} (\bibinfo {year}
  {2014})}\BibitemShut {NoStop}%
\bibitem [{\citenamefont {Joshi}\ \emph {et~al.}(2018)\citenamefont {Joshi},
  \citenamefont {Adli}, \citenamefont {An}, \citenamefont {Clayton},
  \citenamefont {Corde}, \citenamefont {Gessner}, \citenamefont {Hogan},
  \citenamefont {Litos}, \citenamefont {Lu}, \citenamefont {Marsh},
  \citenamefont {Mori}, \citenamefont {Vafaei-Najafabadi}, \citenamefont
  {O'shea}, \citenamefont {Xu}, \citenamefont {White},\ and\ \citenamefont
  {Yakimenko}}]{Joshi2018}%
  \BibitemOpen
  \bibfield  {author} {\bibinfo {author} {\bibfnamefont {C.}~\bibnamefont
  {Joshi}}, \bibinfo {author} {\bibfnamefont {E.}~\bibnamefont {Adli}},
  \bibinfo {author} {\bibfnamefont {W.}~\bibnamefont {An}}, \bibinfo {author}
  {\bibfnamefont {C.~E.}\ \bibnamefont {Clayton}}, \bibinfo {author}
  {\bibfnamefont {S.}~\bibnamefont {Corde}}, \bibinfo {author} {\bibfnamefont
  {S.}~\bibnamefont {Gessner}}, \bibinfo {author} {\bibfnamefont {M.~J.}\
  \bibnamefont {Hogan}}, \bibinfo {author} {\bibfnamefont {M.}~\bibnamefont
  {Litos}}, \bibinfo {author} {\bibfnamefont {W.}~\bibnamefont {Lu}}, \bibinfo
  {author} {\bibfnamefont {K.~A.}\ \bibnamefont {Marsh}}, \bibinfo {author}
  {\bibfnamefont {W.~B.}\ \bibnamefont {Mori}}, \bibinfo {author}
  {\bibfnamefont {N.}~\bibnamefont {Vafaei-Najafabadi}}, \bibinfo {author}
  {\bibfnamefont {B.}~\bibnamefont {O'shea}}, \bibinfo {author} {\bibfnamefont
  {X.}~\bibnamefont {Xu}}, \bibinfo {author} {\bibfnamefont {G.}~\bibnamefont
  {White}}, \ and\ \bibinfo {author} {\bibfnamefont {V.}~\bibnamefont
  {Yakimenko}},\ }\href {\doibase 10.1088/1361-6587/aaa2e3} {\bibfield
  {journal} {\bibinfo  {journal} {Plasma Physics and Controlled Fusion}\
  }\textbf {\bibinfo {volume} {60}},\ \bibinfo {pages} {034001} (\bibinfo
  {year} {2018})}\BibitemShut {NoStop}%
\bibitem [{\citenamefont {Yakimenko}\ \emph {et~al.}(2019)\citenamefont
  {Yakimenko}, \citenamefont {Alsberg}, \citenamefont {Bong}, \citenamefont
  {Bouchard}, \citenamefont {Clarke}, \citenamefont {Emma}, \citenamefont
  {Green}, \citenamefont {Hast}, \citenamefont {Hogan}, \citenamefont
  {Seabury}, \citenamefont {Lipkowitz}, \citenamefont {O'Shea}, \citenamefont
  {Storey}, \citenamefont {White},\ and\ \citenamefont
  {Yocky}}]{Yakimenko2019}%
  \BibitemOpen
  \bibfield  {author} {\bibinfo {author} {\bibfnamefont {V.}~\bibnamefont
  {Yakimenko}}, \bibinfo {author} {\bibfnamefont {L.}~\bibnamefont {Alsberg}},
  \bibinfo {author} {\bibfnamefont {E.}~\bibnamefont {Bong}}, \bibinfo {author}
  {\bibfnamefont {G.}~\bibnamefont {Bouchard}}, \bibinfo {author}
  {\bibfnamefont {C.}~\bibnamefont {Clarke}}, \bibinfo {author} {\bibfnamefont
  {C.}~\bibnamefont {Emma}}, \bibinfo {author} {\bibfnamefont {S.}~\bibnamefont
  {Green}}, \bibinfo {author} {\bibfnamefont {C.}~\bibnamefont {Hast}},
  \bibinfo {author} {\bibfnamefont {M.~J.}\ \bibnamefont {Hogan}}, \bibinfo
  {author} {\bibfnamefont {J.}~\bibnamefont {Seabury}}, \bibinfo {author}
  {\bibfnamefont {N.}~\bibnamefont {Lipkowitz}}, \bibinfo {author}
  {\bibfnamefont {B.}~\bibnamefont {O'Shea}}, \bibinfo {author} {\bibfnamefont
  {D.}~\bibnamefont {Storey}}, \bibinfo {author} {\bibfnamefont
  {G.}~\bibnamefont {White}}, \ and\ \bibinfo {author} {\bibfnamefont
  {G.}~\bibnamefont {Yocky}},\ }\href {\doibase
  10.1103/PhysRevAccelBeams.22.101301} {\bibfield  {journal} {\bibinfo
  {journal} {Physical Review Accelerators and Beams}\ }\textbf {\bibinfo
  {volume} {22}},\ \bibinfo {pages} {101301} (\bibinfo {year}
  {2019})}\BibitemShut {NoStop}%
\bibitem [{\citenamefont {Aschikhin}\ \emph {et~al.}(2016)\citenamefont
  {Aschikhin}, \citenamefont {Behrens}, \citenamefont {Bohlen}, \citenamefont
  {Dale}, \citenamefont {Delbos}, \citenamefont {{di Lucchio}}, \citenamefont
  {Elsen}, \citenamefont {Erbe}, \citenamefont {Felber}, \citenamefont
  {Foster}, \citenamefont {Goldberg}, \citenamefont {Grebenyuk}, \citenamefont
  {Gruse}, \citenamefont {Hidding}, \citenamefont {Hu}, \citenamefont
  {Karstensen}, \citenamefont {Knetsch}, \citenamefont {Kononenko},
  \citenamefont {Libov}, \citenamefont {Ludwig}, \citenamefont {Maier},
  \citenamefont {{Martinez de la Ossa}}, \citenamefont {Mehrling},
  \citenamefont {Palmer}, \citenamefont {Pannek}, \citenamefont {Schaper},
  \citenamefont {Schlarb}, \citenamefont {Schmidt}, \citenamefont {Schreiber},
  \citenamefont {Schwinkendorf}, \citenamefont {Steel}, \citenamefont
  {Streeter}, \citenamefont {Tauscher}, \citenamefont {Wacker}, \citenamefont
  {Weichert}, \citenamefont {Wunderlich}, \citenamefont {Zemella},\ and\
  \citenamefont {Osterhoff}}]{Aschikhin2016}%
  \BibitemOpen
  \bibfield  {author} {\bibinfo {author} {\bibfnamefont {A.}~\bibnamefont
  {Aschikhin}}, \bibinfo {author} {\bibfnamefont {C.}~\bibnamefont {Behrens}},
  \bibinfo {author} {\bibfnamefont {S.}~\bibnamefont {Bohlen}}, \bibinfo
  {author} {\bibfnamefont {J.}~\bibnamefont {Dale}}, \bibinfo {author}
  {\bibfnamefont {N.}~\bibnamefont {Delbos}}, \bibinfo {author} {\bibfnamefont
  {L.}~\bibnamefont {{di Lucchio}}}, \bibinfo {author} {\bibfnamefont
  {E.}~\bibnamefont {Elsen}}, \bibinfo {author} {\bibfnamefont {J.-H.}\
  \bibnamefont {Erbe}}, \bibinfo {author} {\bibfnamefont {M.}~\bibnamefont
  {Felber}}, \bibinfo {author} {\bibfnamefont {B.}~\bibnamefont {Foster}},
  \bibinfo {author} {\bibfnamefont {L.}~\bibnamefont {Goldberg}}, \bibinfo
  {author} {\bibfnamefont {J.}~\bibnamefont {Grebenyuk}}, \bibinfo {author}
  {\bibfnamefont {J.-N.}\ \bibnamefont {Gruse}}, \bibinfo {author}
  {\bibfnamefont {B.}~\bibnamefont {Hidding}}, \bibinfo {author} {\bibfnamefont
  {Z.}~\bibnamefont {Hu}}, \bibinfo {author} {\bibfnamefont {S.}~\bibnamefont
  {Karstensen}}, \bibinfo {author} {\bibfnamefont {A.}~\bibnamefont {Knetsch}},
  \bibinfo {author} {\bibfnamefont {O.}~\bibnamefont {Kononenko}}, \bibinfo
  {author} {\bibfnamefont {V.}~\bibnamefont {Libov}}, \bibinfo {author}
  {\bibfnamefont {K.}~\bibnamefont {Ludwig}}, \bibinfo {author} {\bibfnamefont
  {A.~R.}\ \bibnamefont {Maier}}, \bibinfo {author} {\bibfnamefont
  {A.}~\bibnamefont {{Martinez de la Ossa}}}, \bibinfo {author} {\bibfnamefont
  {T.}~\bibnamefont {Mehrling}}, \bibinfo {author} {\bibfnamefont {C.~A.~J.}\
  \bibnamefont {Palmer}}, \bibinfo {author} {\bibfnamefont {F.}~\bibnamefont
  {Pannek}}, \bibinfo {author} {\bibfnamefont {L.}~\bibnamefont {Schaper}},
  \bibinfo {author} {\bibfnamefont {H.}~\bibnamefont {Schlarb}}, \bibinfo
  {author} {\bibfnamefont {B.}~\bibnamefont {Schmidt}}, \bibinfo {author}
  {\bibfnamefont {S.}~\bibnamefont {Schreiber}}, \bibinfo {author}
  {\bibfnamefont {J.-P.}\ \bibnamefont {Schwinkendorf}}, \bibinfo {author}
  {\bibfnamefont {H.}~\bibnamefont {Steel}}, \bibinfo {author} {\bibfnamefont
  {M.}~\bibnamefont {Streeter}}, \bibinfo {author} {\bibfnamefont
  {G.}~\bibnamefont {Tauscher}}, \bibinfo {author} {\bibfnamefont
  {V.}~\bibnamefont {Wacker}}, \bibinfo {author} {\bibfnamefont
  {S.}~\bibnamefont {Weichert}}, \bibinfo {author} {\bibfnamefont
  {S.}~\bibnamefont {Wunderlich}}, \bibinfo {author} {\bibfnamefont
  {J.}~\bibnamefont {Zemella}}, \ and\ \bibinfo {author} {\bibfnamefont
  {J.}~\bibnamefont {Osterhoff}},\ }\href {\doibase 10.1016/j.nima.2015.10.005}
  {\bibfield  {journal} {\bibinfo  {journal} {Nuclear Instruments and Methods
  in Physics Research, Section A: Accelerators, Spectrometers, Detectors and
  Associated Equipment}\ }\textbf {\bibinfo {volume} {806}},\ \bibinfo {pages}
  {175} (\bibinfo {year} {2016})}\BibitemShut {NoStop}%
\bibitem [{\citenamefont {Lindstr{\o}m}\ \emph {et~al.}(2021)\citenamefont
  {Lindstr{\o}m}, \citenamefont {Garland}, \citenamefont {Schr{\"{o}}der},
  \citenamefont {Boulton}, \citenamefont {Boyle}, \citenamefont {Chappell},
  \citenamefont {D'Arcy}, \citenamefont {Gonzalez}, \citenamefont {Knetsch},
  \citenamefont {Libov}, \citenamefont {Loisch}, \citenamefont {{Martinez de la
  Ossa}}, \citenamefont {Niknejadi}, \citenamefont {P{\~{o}}der}, \citenamefont
  {Schaper}, \citenamefont {Schmidt}, \citenamefont {Sheeran}, \citenamefont
  {Wesch}, \citenamefont {Wood},\ and\ \citenamefont
  {Osterhoff}}]{Lindstrom2021a}%
  \BibitemOpen
  \bibfield  {author} {\bibinfo {author} {\bibfnamefont {C.~A.}\ \bibnamefont
  {Lindstr{\o}m}}, \bibinfo {author} {\bibfnamefont {J.~M.}\ \bibnamefont
  {Garland}}, \bibinfo {author} {\bibfnamefont {S.}~\bibnamefont
  {Schr{\"{o}}der}}, \bibinfo {author} {\bibfnamefont {L.}~\bibnamefont
  {Boulton}}, \bibinfo {author} {\bibfnamefont {G.}~\bibnamefont {Boyle}},
  \bibinfo {author} {\bibfnamefont {J.}~\bibnamefont {Chappell}}, \bibinfo
  {author} {\bibfnamefont {R.}~\bibnamefont {D'Arcy}}, \bibinfo {author}
  {\bibfnamefont {P.}~\bibnamefont {Gonzalez}}, \bibinfo {author}
  {\bibfnamefont {A.}~\bibnamefont {Knetsch}}, \bibinfo {author} {\bibfnamefont
  {V.}~\bibnamefont {Libov}}, \bibinfo {author} {\bibfnamefont
  {G.}~\bibnamefont {Loisch}}, \bibinfo {author} {\bibfnamefont
  {A.}~\bibnamefont {{Martinez de la Ossa}}}, \bibinfo {author} {\bibfnamefont
  {P.}~\bibnamefont {Niknejadi}}, \bibinfo {author} {\bibfnamefont
  {K.}~\bibnamefont {P{\~{o}}der}}, \bibinfo {author} {\bibfnamefont
  {L.}~\bibnamefont {Schaper}}, \bibinfo {author} {\bibfnamefont
  {B.}~\bibnamefont {Schmidt}}, \bibinfo {author} {\bibfnamefont
  {B.}~\bibnamefont {Sheeran}}, \bibinfo {author} {\bibfnamefont
  {S.}~\bibnamefont {Wesch}}, \bibinfo {author} {\bibfnamefont
  {J.}~\bibnamefont {Wood}}, \ and\ \bibinfo {author} {\bibfnamefont
  {J.}~\bibnamefont {Osterhoff}},\ }\href {\doibase
  10.1103/PhysRevLett.126.014801} {\bibfield  {journal} {\bibinfo  {journal}
  {Physical Review Letters}\ }\textbf {\bibinfo {volume} {126}},\ \bibinfo
  {pages} {014801} (\bibinfo {year} {2021})}\BibitemShut {NoStop}%
\bibitem [{\citenamefont {Floettmann}(2003)}]{Floettmann2003}%
  \BibitemOpen
  \bibfield  {author} {\bibinfo {author} {\bibfnamefont {K.}~\bibnamefont
  {Floettmann}},\ }\href {\doibase 10.1103/PhysRevSTAB.6.034202} {\bibfield
  {journal} {\bibinfo  {journal} {Physical Review Special Topics - Accelerators
  and Beams}\ }\textbf {\bibinfo {volume} {6}},\ \bibinfo {pages} {034202}
  (\bibinfo {year} {2003})}\BibitemShut {NoStop}%
\bibitem [{\citenamefont {Lindstr{\o}m}\ and\ \citenamefont
  {Adli}(2016)}]{Lindstrom2016}%
  \BibitemOpen
  \bibfield  {author} {\bibinfo {author} {\bibfnamefont {C.~A.}\ \bibnamefont
  {Lindstr{\o}m}}\ and\ \bibinfo {author} {\bibfnamefont {E.}~\bibnamefont
  {Adli}},\ }\href {\doibase 10.1103/PhysRevAccelBeams.19.071002} {\bibfield
  {journal} {\bibinfo  {journal} {Physical Review Accelerators and Beams}\
  }\textbf {\bibinfo {volume} {19}},\ \bibinfo {pages} {071002} (\bibinfo
  {year} {2016})}\BibitemShut {NoStop}%
\bibitem [{\citenamefont {Tzoufras}\ \emph {et~al.}(2008)\citenamefont
  {Tzoufras}, \citenamefont {Lu}, \citenamefont {Tsung}, \citenamefont {Huang},
  \citenamefont {Mori}, \citenamefont {Katsouleas}, \citenamefont {Vieira},
  \citenamefont {Fonseca},\ and\ \citenamefont {Silva}}]{Tzoufras2008}%
  \BibitemOpen
  \bibfield  {author} {\bibinfo {author} {\bibfnamefont {M.}~\bibnamefont
  {Tzoufras}}, \bibinfo {author} {\bibfnamefont {W.}~\bibnamefont {Lu}},
  \bibinfo {author} {\bibfnamefont {F.~S.}\ \bibnamefont {Tsung}}, \bibinfo
  {author} {\bibfnamefont {C.}~\bibnamefont {Huang}}, \bibinfo {author}
  {\bibfnamefont {W.~B.}\ \bibnamefont {Mori}}, \bibinfo {author}
  {\bibfnamefont {T.}~\bibnamefont {Katsouleas}}, \bibinfo {author}
  {\bibfnamefont {J.}~\bibnamefont {Vieira}}, \bibinfo {author} {\bibfnamefont
  {R.~A.}\ \bibnamefont {Fonseca}}, \ and\ \bibinfo {author} {\bibfnamefont
  {L.~O.}\ \bibnamefont {Silva}},\ }\href {\doibase
  10.1103/PhysRevLett.101.145002} {\bibfield  {journal} {\bibinfo  {journal}
  {Physical Review Letters}\ }\textbf {\bibinfo {volume} {101}},\ \bibinfo
  {pages} {145002} (\bibinfo {year} {2008})},\ \Eprint
  {http://arxiv.org/abs/0809.0227} {arXiv:0809.0227} \BibitemShut {NoStop}%
\bibitem [{\citenamefont {Nieter}\ and\ \citenamefont
  {Cary}(2004)}]{nieter:2004jcp}%
  \BibitemOpen
  \bibfield  {author} {\bibinfo {author} {\bibfnamefont {C.}~\bibnamefont
  {Nieter}}\ and\ \bibinfo {author} {\bibfnamefont {J.~R.}\ \bibnamefont
  {Cary}},\ }\href {\doibase 10.1016/j.jcp.2003.11.004} {\bibfield  {journal}
  {\bibinfo  {journal} {Journal of Computational Physics}\ }\textbf {\bibinfo
  {volume} {196}},\ \bibinfo {pages} {448} (\bibinfo {year}
  {2004})}\BibitemShut {NoStop}%
\bibitem [{\citenamefont {Litos}\ \emph {et~al.}(2019)\citenamefont {Litos},
  \citenamefont {Ariniello}, \citenamefont {Doss}, \citenamefont {Hunt-Stone},\
  and\ \citenamefont {Cary}}]{Litos2019}%
  \BibitemOpen
  \bibfield  {author} {\bibinfo {author} {\bibfnamefont {M.~D.}\ \bibnamefont
  {Litos}}, \bibinfo {author} {\bibfnamefont {R.}~\bibnamefont {Ariniello}},
  \bibinfo {author} {\bibfnamefont {C.~E.}\ \bibnamefont {Doss}}, \bibinfo
  {author} {\bibfnamefont {K.}~\bibnamefont {Hunt-Stone}}, \ and\ \bibinfo
  {author} {\bibfnamefont {J.~R.}\ \bibnamefont {Cary}},\ }\href {\doibase
  10.1098/rsta.2018.0181} {\bibfield  {journal} {\bibinfo  {journal}
  {Philosophical Transactions of the Royal Society A: Mathematical, Physical
  and Engineering Sciences}\ }\textbf {\bibinfo {volume} {377}},\ \bibinfo
  {pages} {20180181} (\bibinfo {year} {2019})}\BibitemShut {NoStop}%
\bibitem [{\citenamefont {Pak}\ \emph {et~al.}(2010)\citenamefont {Pak},
  \citenamefont {Marsh}, \citenamefont {Martins}, \citenamefont {Lu},
  \citenamefont {Mori},\ and\ \citenamefont {Joshi}}]{Pak2010}%
  \BibitemOpen
  \bibfield  {author} {\bibinfo {author} {\bibfnamefont {A.}~\bibnamefont
  {Pak}}, \bibinfo {author} {\bibfnamefont {K.~A.}\ \bibnamefont {Marsh}},
  \bibinfo {author} {\bibfnamefont {S.~F.}\ \bibnamefont {Martins}}, \bibinfo
  {author} {\bibfnamefont {W.}~\bibnamefont {Lu}}, \bibinfo {author}
  {\bibfnamefont {W.~B.}\ \bibnamefont {Mori}}, \ and\ \bibinfo {author}
  {\bibfnamefont {C.}~\bibnamefont {Joshi}},\ }\href {\doibase
  10.1103/PhysRevLett.104.025003} {\bibfield  {journal} {\bibinfo  {journal}
  {Physical Review Letters}\ }\textbf {\bibinfo {volume} {104}},\ \bibinfo
  {pages} {025003} (\bibinfo {year} {2010})}\BibitemShut {NoStop}%
\bibitem [{\citenamefont {Hidding}\ \emph {et~al.}(2012)\citenamefont
  {Hidding}, \citenamefont {Pretzler}, \citenamefont {Rosenzweig},
  \citenamefont {K{\"{o}}nigstein}, \citenamefont {Schiller},\ and\
  \citenamefont {Bruhwiler}}]{Hidding2012}%
  \BibitemOpen
  \bibfield  {author} {\bibinfo {author} {\bibfnamefont {B.}~\bibnamefont
  {Hidding}}, \bibinfo {author} {\bibfnamefont {G.}~\bibnamefont {Pretzler}},
  \bibinfo {author} {\bibfnamefont {J.~B.}\ \bibnamefont {Rosenzweig}},
  \bibinfo {author} {\bibfnamefont {T.}~\bibnamefont {K{\"{o}}nigstein}},
  \bibinfo {author} {\bibfnamefont {D.}~\bibnamefont {Schiller}}, \ and\
  \bibinfo {author} {\bibfnamefont {D.~L.}\ \bibnamefont {Bruhwiler}},\ }\href
  {\doibase 10.1103/PhysRevLett.108.035001} {\bibfield  {journal} {\bibinfo
  {journal} {Physical Review Letters}\ }\textbf {\bibinfo {volume} {108}},\
  \bibinfo {pages} {035001} (\bibinfo {year} {2012})}\BibitemShut {NoStop}%
\end{thebibliography}%

\end{document}